\documentclass[aps,prc,preprint,superscriptaddress,showpacs]{revtex4}
\usepackage[dvips]{graphicx}
\usepackage{amsmath,amssymb,times}

\newcommand{\bea}{\begin{eqnarray}}
\newcommand{\eea}{\end{eqnarray}}

\renewcommand{\a}{\alpha}

\newcommand{\la}{\left\langle}

\def\la{\mathrel{\mathpalette\fun <}}

\def\fun#1#2{\lower3.6pt\vbox{\baselineskip0pt\lineskip.9pt
  \ialign{$\mathsurround=0pt#1\hfil##\hfil$\crcr#2\crcr\sim\crcr}}}

\DeclareSymbolFont{boldletters}{OML}{cmm} {b}{it}
\DeclareSymbolFontAlphabet{\mathbit}{boldletters}
\DeclareMathSymbol{\alpha}{\mathalpha}{letters}{"0B}
\DeclareMathSymbol{\beta}{\mathalpha}{letters}{"0C}
\DeclareMathSymbol{\gamma}{\mathalpha}{letters}{"0D}
\DeclareMathSymbol{\delta}{\mathalpha}{letters}{"0E}
\DeclareMathSymbol{\epsilon}{\mathalpha}{letters}{"0F}
\DeclareMathSymbol{\zeta}{\mathalpha}{letters}{"10}
\DeclareMathSymbol{\eta}{\mathalpha}{letters}{"11}
\DeclareMathSymbol{\theta}{\mathalpha}{letters}{"12}
\DeclareMathSymbol{\iota}{\mathalpha}{letters}{"13}
\DeclareMathSymbol{\kappa}{\mathalpha}{letters}{"14}
\DeclareMathSymbol{\lambda}{\mathalpha}{letters}{"15}
\DeclareMathSymbol{\mu}{\mathalpha}{letters}{"16}
\DeclareMathSymbol{\nu}{\mathalpha}{letters}{"17}
\DeclareMathSymbol{\xi}{\mathalpha}{letters}{"18}
\DeclareMathSymbol{\pi}{\mathalpha}{letters}{"19}
\DeclareMathSymbol{\rho}{\mathalpha}{letters}{"1A}
\DeclareMathSymbol{\sigma}{\mathalpha}{letters}{"1B}
\DeclareMathSymbol{\tau}{\mathalpha}{letters}{"1C}
\DeclareMathSymbol{\upsilon}{\mathalpha}{letters}{"1D}
\DeclareMathSymbol{\phi}{\mathalpha}{letters}{"1E}
\DeclareMathSymbol{\chi}{\mathalpha}{letters}{"1F}
\DeclareMathSymbol{\psi}{\mathalpha}{letters}{"20}
\DeclareMathSymbol{\omega}{\mathalpha}{letters}{"21}
\DeclareMathSymbol{\varepsilon}{\mathalpha}{letters}{"22}
\DeclareMathSymbol{\vartheta}{\mathalpha}{letters}{"23}
\DeclareMathSymbol{\varpi}{\mathalpha}{letters}{"24}
\DeclareMathSymbol{\varrho}{\mathalpha}{letters}{"25}
\DeclareMathSymbol{\varsigma}{\mathalpha}{letters}{"26}
\DeclareMathSymbol{\varphi}{\mathalpha}{letters}{"27}
\DeclareMathSymbol{\Gamma}{\mathalpha}{letters}{"00}
\DeclareMathSymbol{\Delta}{\mathalpha}{letters}{"01}
\DeclareMathSymbol{\Theta}{\mathalpha}{letters}{"02}
\DeclareMathSymbol{\Lambda}{\mathalpha}{letters}{"03}
\DeclareMathSymbol{\Xi}{\mathalpha}{letters}{"04}
\DeclareMathSymbol{\Pi}{\mathalpha}{letters}{"05}
\DeclareMathSymbol{\Sigma}{\mathalpha}{letters}{"06}
\DeclareMathSymbol{\Upsilon}{\mathalpha}{letters}{"07}
\DeclareMathSymbol{\Phi}{\mathalpha}{letters}{"08}
\DeclareMathSymbol{\Psi}{\mathalpha}{letters}{"09}
\DeclareMathSymbol{\Omega}{\mathalpha}{letters}{"0A}

\begin{document}
\preprint{SAGA-HE-255}
\title{QCD phase diagram at imaginary baryon 
and isospin chemical potentials}

\author{Yuji Sakai}
\email[]{sakai@phys.kyushu-u.ac.jp}
\affiliation{Department of Physics, Graduate School of Sciences, Kyushu University,
             Fukuoka 812-8581, Japan}

\author{Hiroaki Kouno}
\email[]{kounoh@cc.saga-u.ac.jp}
\affiliation{Department of Physics, Saga University,
             Saga 840-8502, Japan}

\author{Masanobu Yahiro}
\email[]{yahiro@phys.kyushu-u.ac.jp}
\affiliation{Department of Physics, Graduate School of Sciences, Kyushu University,
             Fukuoka 812-8581, Japan}

\date{\today}

\begin{abstract}
We explore the phase diagram of two-flavor QCD at imaginary values of 
baryon and isospin chemical potentials, $\mu_{\rm B}$ and $\mu_{\rm iso}$, 
analyzing the thermodynamic potential of QCD analytically 
and that of the Polyakov-loop extended Nambu--Jona-Lasinio (PNJL) model 
numerically. 
QCD has no pion condensation at imaginary $\mu_{\rm B}$ and $\mu_{\rm iso}$, 
and therefore has discrete symmetries that are not present at real 
$\mu_{\rm B}$ and $\mu_{\rm iso}$. 
The PNJL model possesses all the discrete symmetries. 
The PNJL model can reproduce qualitatively lattice QCD data 
presented very lately. 
\end{abstract}

\maketitle

\section{Introduction}
\label{Introduction}

Quantum Chromodynamics (QCD) as a fundamental theory on strong interaction is 
well defined, since it is renormalizable and parameter free. 
However, thermodynamics of QCD is 
not well understood because of its nonperturbative nature. 
In particular, QCD phase diagram is essential for 
understanding not only natural 
phenomena such as compact stars and the early universe but also 
laboratory experiments such as relativistic heavy-ion collisions. 
Quantitative calculations of the phase diagram from 
first-principle lattice QCD (LQCD) have the well known sign problem 
when the baryon chemical potential ($\mu_{\rm B}$) is real; 
for example, see Ref.~\cite{Kogut} and references therein. 
For later convenience, we use the quark-number chemical potential 
$\mu_{\rm q}=\mu_{\rm B}/3$ instead of $\mu_{\rm B}$.  
So far, several approaches have been proposed to circumvent the difficulty; 
for example, the reweighting method~\cite{Fodor}, 
the Taylor expansion method~\cite{Allton} and 
the analytic continuation from imaginary $\mu_{\rm q}$ 
to real $\mu_{\rm q}$~\cite{FP,D'Elia-1,Chen}. 
However, those are still far from perfection.  

As an approach complementary to first-principle LQCD, 
we can consider effective models such as the  
Nambu--Jona-Lasinio (NJL) model~
\cite{NJ1,AY,KKKN,Kashiwa1,He} and 
the Polyakov-loop extended Nambu--Jona-Lasinio (PNJL) 
model~\cite{Meisinger,Fukushima,Fukushima2,Ghos,Megias,Ratti,Ciminale,
Rossner,Hansen,Sasaki,Schaefer,Zhang,Mukherjee,Kashiwa2,Fu,Abuki,AF,Hell,
Sakai1,Sakai2,Kashiwa3,Kouno}. 
The NJL model describes the chiral symmetry breaking, but not 
the confinement mechanism. 
The PNJL model is constructed so as to treat both the Polyakov loop and 
the chiral symmetry breaking~\cite{Fukushima}.

In the NJL-type models, 
the input parameters are determined at $\mu_{\rm q}=0$ and $T \ge 0$, 
where $T$ is temperature. 
It is then highly nontrivial whether the models predict properly dynamics 
of QCD at finite $\mu_{\rm q}$.  
This should be tested from QCD. 
Fortunately, this is possible at imaginary $\mu_{\rm q}$, 
since LQCD has no sign problem there.

Roberge and Weiss found~\cite{RW} that the thermodynamic potential 
$\Omega_{\rm QCD}(\theta_{\rm q})$ 
of QCD at imaginary chemical potential $\mu_{\rm q}=iT \theta_{\rm q} $
has a periodicity 
$\Omega_{\rm QCD}(\theta_{\rm q})=\Omega_{\rm QCD}(\theta_{\rm q}+2\pi k/3)$, 
showing that $\Omega_{\rm QCD}(\theta_{\rm q}+2\pi k/3)$ is 
transformed into $\Omega_{\rm QCD}(\theta_{\rm q})$ by 
the ${\mathbb Z}_3$ transformation with integer $k$. 
This means that QCD is invariant under 
a combination of the ${\mathbb Z}_3$ transformation and 
a parameter transformation $\theta_{\rm q} 
\to \theta_{\rm q}+2k\pi/3$~\cite{Sakai1}, 
\begin{eqnarray}
q \to Uq, \quad 
A_{\nu} \to UA_{\nu}U^{-1} - i/g (\partial_{\nu}U)U^{-1}, \quad 
\theta_{\rm q} \to \theta_{\rm q}+2\pi k/3, 
\label{ez3}
\end{eqnarray}
where $U(x,\tau)$ are elements of SU(3) with 
$
U(x,\beta=1/T)=\exp(-2i \pi k/3)U(x,0)   
$
and $q$ is the quark field. 
We call this combination the extended 
${\mathbb Z}_3$ transformation. 
Thus, $\Omega_{\rm QCD}(\theta_{\rm q})$ 
has the extended ${\mathbb Z}_3$ symmetry, and hence 
quantities invariant under the extended ${\mathbb Z}_3$ transformation
have the RW periodicity~\cite{Sakai1}.

At the present stage, the PNJL model is 
only a realistic effective model that possesses both 
the extended ${\mathbb Z}_3$ symmetry and 
chiral symmetry~\cite{Sakai1}. 
This property guarantees that the phase diagram evaluated by the 
PNJL model has the RW periodicity in the imaginary $\mu_{\rm q}$ region, and 
therefore makes it possible to compare the PNJL result 
with LQCD data~\cite{FP,D'Elia-1,Chen} 
quantitatively in the imaginary $\mu_{\rm q}$ region. 
Actually, the PNJL model succeeds in reproducing the LQCD data by 
introducing the 
vector-type four-quark interaction~\cite{AY,KKKN,Kashiwa1} and the scalar-type 
eight-quark interaction~\cite{Kashiwa1}.  
The QCD phase diagram in the real $\mu_{\rm q}$ region is predicted by 
the PNJL model with the parameter set~\cite{Sakai2} that reproduces 
the LQCD data at imaginary $\mu_{\rm q}$. 
The critical endpoint can survive, even if 
the vector-type four-quark interaction is taken into account.

LQCD has no sign problem also at finite isospin chemical potential 
($\mu_{\rm iso}$)~\cite{Son-Stephanov}. 
This is true for both real and imaginary isospin chemical potentials, 
as explicitly shown in Sec.~\ref{Symmetries}. For later convenience, 
we use the ``modified" 
isospin chemical potential $\mu_{\rm I}=\mu_{\rm iso}/2$ 
instead of $\mu_{\rm iso}$ itself. 
Very recently, 
LQCD data were 
measured at both real and imaginary $\mu_{\rm I}$~\cite{Cea} 
and also in the case that both $\mu_{\rm I}$ and $\mu_{\rm q}$ are 
imaginary~\cite{D'Elia-2}. 
The PNJL model has already 
been applied to the real $\mu_{\rm I}$ case~\cite{Zhang,Mukherjee}, 
but not to the imaginary $\mu_{\rm I}$ case.

In this paper, we explore the phase diagram of two-flavor QCD
at pure imaginary values of $\mu_{\rm q}$ and $\mu_{\rm I}$, 
by analyzing the partition function of QCD analytically and 
the thermodynamic potential of PNJL numerically. 
As the primary result, we will show that 
the pion condensation does not occur at imaginary 
$\mu_{\rm I}$ and $\mu_{\rm q}$ and hence isospin and baryon number are 
conserved. As a consequence of this property, 
$\Omega_{\rm QCD}$ 
has higher discrete symmetries at imaginary 
$\mu_{\rm I}$ and $\mu_{\rm q}$ than at 
real $\mu_{\rm I}$ and $\mu_{\rm q}$. The PNJL model possesses all the 
symmetries, and then the model reproduces LQCD data~\cite{Cea,D'Elia-2} 
qualitatively at imaginary $\mu_{\rm I}$ and $\mu_{\rm q}$. 
Finally, the phase diagram at imaginary $\mu_{\rm I}$ and $\mu_{\rm q}$ 
is predicted by the PNJL model.

In Sec.~\ref{Symmetries}, it is shown 
at imaginary $\mu_{\rm iso}$ and $\mu_{\rm q}$ that 
no pion condensation takes place and then QCD 
has some discrete symmetries. 
A simple explanation of the PNJL model is made in Sec.~\ref{PNJL}, 
and numerical results of PNJL calculations are presented 
in Sec.~\ref{Numerical-results}. 
Section~\ref{Summary} is devoted to summary.

\section{Discrete symmetries of QCD}
\label{Symmetries}

Roberge and Weiss showed the RW periodicity in the one-flavor case~\cite{RW}, 
assuming that baryon number is conserved. 
Extending their proof to the two-flavor case, we will prove that 
$\Omega_{\rm QCD}(\theta_{\rm q},\theta_{\rm I})$ has some 
discrete symmetries 
at imaginary $\mu_{\rm q}$ and $\mu_{\rm I}$. 
In this proof, we first assume that baryon number and isospin, 
i.e., $u$-quark and $d$-quark numbers, are conserved, but this assumption is 
confirmed to be true at the end of this section.

The thermodynamic potential 
$\Omega_{\rm QCD}(\theta_{\rm q},\theta_{\rm I})$ (per unit volume) 
is related to the partition function $Z(\theta_{\rm q},\theta_{\rm I})$ 
as $\Omega_{\rm QCD}=-T\ln(Z)/V$, where $V$ represents the 
infinite volume we are thinking. 
The functional integral form of 
$Z$ in Euclidean spacetime with time interval $\tau \in (0,\beta=1/T)$ is 
\begin{eqnarray}
&Z=\int Dq D\bar{q} DA 
\exp\left[ - S \right] , \nonumber \\
&S=\int d^4 x 
{\left[\bar{q}
(\gamma_{\nu} D_{\nu} - \gamma_4{\hat \mu} + \hat{m}_0)q 
+{1\over{4}}F_{\mu\nu}^2 \right]} ,
\label{eq:EQ1}
\end{eqnarray}
where $q=(q_u, q_d)^{\rm T}$ is the two-flavor quark field, 
${\hat m_0}={\rm diag}(m_u, m_d)$ is the current quark mass, and 
$D_{\nu}$ is the covariant derivative. 
We take the isospin symmetric limit of $m_u=m_d=m_0$.

The chemical potential matrix ${\hat \mu}$ is 
defined by ${\hat \mu}={\rm diag}(\mu_u, \mu_d)$ with 
the $u$-quark number chemical potential ($\mu_{u}$) and 
the $d$-quark one ($\mu_{d}$). This is 
equivalent to introducing the baryon and isospin chemical potentials, 
$\mu_{\rm B}$ and $\mu_{\rm iso}$, 
coupled respectively to the baryon charge ${\bar B}$ and to the isospin 
charge ${\bar I_3}$:
\begin{align}
{\hat \mu}=\mu_{\rm q} \tau_0 + \mu_{\rm I} \tau_3 
\end{align}
with 
\begin{align}
\mu_{\rm q}=\frac{\mu_{u}+\mu_{d}}{2}=\frac{\mu_{\rm B}}{3},
~~\mu_{\rm I}=\frac{\mu_{u}-\mu_{d}}{2}=\frac{\mu_{\rm iso}}{2} , 
\end{align}
where $\tau_0$ is the unit matrix and $\tau_i$ ($i=1, 2, 3$) 
are the Pauli matrices in flavor space. 
Note that $\mu_{\rm I}$ is half the isospin chemical potential 
($\mu_{\rm iso}$). 
For later convenience, the dimensionless chemical potentials, 
$\theta_y$ ($y=u, d, {\rm q}, {\rm I}$), are introduced by $\mu_y=iT\theta_y$.

Now, we transform the quark field $q$ as 
\begin{align}
q\to(\exp[i\theta_u \tau/\beta]q_u, \exp[i\theta_d \tau/\beta]q_d)^T
=\exp[i\theta_{\rm q}\tau/\beta]
\bigl( \cos{[\theta_{\rm I} \tau/\beta}] \tau_0 + 
i\sin[\theta_{\rm I} \tau/\beta] \tau_3 \bigr) q . 
\label{1st-trans}
\end{align}
This transformation leads $Z$ to 
\begin{eqnarray}
&Z = \int Dq D\bar{q} DA \exp\left[-S\right], \notag \\
&S = \int d^4 x {\left[\bar{q}(\gamma_{\nu}D_{\nu}+\hat{m}_0)q 
+\frac{1}{4}F_{\mu\nu}^2 \right]} 
\label{eq:EQ2}
\end{eqnarray}
with the boundary conditions 
\begin{eqnarray}
&q_{u}(x,\beta)=-\exp[i(\theta_{\rm q}+\theta_{\rm I})]q_{u}(x,0) , 
\notag \\
&q_{d}(x,\beta)=-\exp[i(\theta_{\rm q}-\theta_{\rm I})]q_{d}(x,0) .
\label{BC1}
\end{eqnarray}

Under the ${\mathbb Z}_{3}$ transformation, i.e., the first and second 
transformations of \eqref{ez3}, $Z$ keeps the same form as \eqref{eq:EQ2}, 
but the boundary conditions are changed into 
\begin{eqnarray}
&q_{u}(x,\beta)=-\exp[i(\theta_{\rm q}+\theta_{\rm I}-2\pi k/3)]q_{u}(x,0) , 
\notag \\
&q_{d}(x,\beta)=-\exp[i(\theta_{\rm q}-\theta_{\rm I}-2\pi k/3)]q_{d}(x,0) .
\label{BC2}
\end{eqnarray}
The functional form of \eqref{eq:EQ2} with the boundary conditions 
\eqref{BC2} means $Z(\theta_{\rm q}-2\pi k/3,\theta_{\rm I})$. 
Since the ${\mathbb Z}_{3}$ transformation corresponds to 
the redefinition of fields in the path integration, 
we can reach the equality 
\begin{eqnarray}
Z(\theta_{\rm q},\theta_{\rm I})=Z(\theta_{\rm q}-2\pi k/3,\theta_{\rm I}).
\label{RW-periodicity}
\end{eqnarray}
Further, using \eqref{BC1}, one can see that 
\begin{eqnarray}
&Z(\theta_{\rm q},\theta_{\rm I})=Z(\theta_{\rm q},\theta_{\rm I}+2\pi) ,
\label{2pi-periodicity}
\\
&Z(\theta_{\rm q}+\pi,\theta_{\rm I})=Z(\theta_{\rm q},\theta_{\rm I}+\pi).  
\label{pi-periodicity}
\end{eqnarray}

In the isospin symmetric limit $m_u=m_d$, $Z$ is invariant under 
the interchange $u \leftrightarrow d$. This means that
\begin{eqnarray}
Z(\theta_{\rm q},\theta_{\rm I})=Z(\theta_{\rm q},-\theta_{\rm I}).
\label{even-I}
\end{eqnarray}
Furthermore, $Z$ is invariant under charge conjugation, 
when $\theta_{\rm q}$ and $\theta_{\rm I}$ are transformed as 
$\theta_{\rm q} \to -\theta_{\rm q}$ and $\theta_{\rm I} \to -\theta_{\rm I}$. 
This indicates that 
\begin{eqnarray}
Z(\theta_{\rm q},\theta_{\rm I})=Z(-\theta_{\rm q},-\theta_{\rm I}).
\label{even-qI}
\end{eqnarray}
Equations \eqref{even-I} and \eqref{even-qI} show that 
\begin{eqnarray}
Z(\theta_{\rm q},\theta_{\rm I})=Z(-\theta_{\rm q},\theta_{\rm I}).
\label{even-q}
\end{eqnarray}
Thus, $Z$ is $\theta_{\rm q}$-even and $\theta_{\rm I}$-even. 
The relations \eqref{RW-periodicity}, \eqref{pi-periodicity}, 
\eqref{even-I} and \eqref{even-q} lead to new ones 
\begin{eqnarray}
Z(\theta_{\rm q},\pi\pm\theta_{\rm I})=Z(\theta_{\rm q}+\pi,\pm\theta_{\rm I})
&=&Z(\theta_{\rm q}+\pi/3,\theta_{\rm I}), ~~~\quad \\
\label{pi}
Z(\pi/3-\theta_{\rm q}, \theta_{\rm I})=Z(\theta_{\rm q}-\pi/3, \theta_{\rm I})
&=&Z(\theta_{\rm q}+\pi, \theta_{\rm I})
=Z(\theta_{\rm q}, \theta_{\rm I}+\pi). 
\label{pi/3}
\end{eqnarray}
The thermodynamic potential $\Omega_{\rm QCD}=-T\ln(Z)/V$  and 
the chiral condensate $\sigma=d\Omega_{\rm QCD}/dm_{0}$
have the same symmetries as $Z$ in \eqref{RW-periodicity}-\eqref{pi/3}.

Making the fermionic path integration in (\ref{eq:EQ1}), one can get 
the determinant $\det\Delta$ with 
$\Delta=\gamma_{\nu}D_{\nu}  - \gamma_{4}{\hat \mu} + {\hat m}_0$. 
This determinant is real, since ${\hat \mu}^{*}=-{\hat \mu}$ 
and then~\cite{Son-Stephanov} 
\begin{eqnarray}
(\det\Delta)^*=\det\Delta^{\dagger}
=\det(\gamma_{5}\Delta\gamma_{5})=\det\Delta \;.
\end{eqnarray}
Further, $\Delta$ has an explicit form of 
\begin{eqnarray}
\det\Delta &=& \det \left[ m_0^2 I + 
(\sigma \cdot D-\mu_u I)^{\dagger}(\sigma \cdot D-\mu_u I) \right] 
\notag \\ 
&\times& \det \left[ m_0^2 I + 
(\sigma \cdot D-\mu_d I)^{\dagger}(\sigma \cdot D-\mu_d I) \right] , \quad
\label{delta-1}
\end{eqnarray}
where $I$ is the $2 \times 2$ unit matrix and 
$\sigma \cdot D=ID_4+i{\vec \sigma} \cdot {\vec D}$. 
Each of the first and second determinants 
on the right-hand side of \eqref{delta-1} 
is the square of a real number. Hence, $\det\Delta$ is 
positive in the case (i) that both $\mu_{\rm q}$ 
and $\mu_{\rm I}$ are imaginary.

Similarly, in the case (ii) that 
$\mu_{\rm q}$ is imaginary and $\mu_{\rm I}$ is real, 
${\hat \mu}$ satisfies ${\hat \mu}^{*}=-\tau_1{\hat \mu}\tau_1$  
and then~\cite{Son-Stephanov}  
\begin{eqnarray}
(\det\Delta)^*=\det\Delta^{\dagger}
=\det(\gamma_{5}\tau_1\Delta\tau_1\gamma_{5})=\det\Delta \;.
\end{eqnarray}
This shows that $\det\Delta$ is real. Furthermore, the determinant is given by 
\begin{eqnarray}
\det\Delta &=& 
\det \left[ m_0^2 I + 
(\sigma \cdot D-\mu_d I)^{\dagger}(\sigma \cdot D-\mu_u I) \right] 
\notag \\ 
&\times& \det \left[ m_0^2 I + 
(\sigma \cdot D-\mu_u I)^{\dagger}(\sigma \cdot D-\mu_d I) \right] . \quad 
\end{eqnarray}
This determinant is also the square of a real number and then positive. 
Thus, in both cases of (i) and (ii), LQCD has no sign problem.

The Polyakov loop $\hat{\Phi}$ and its Hermitian conjugate 
$\hat{\Phi}^{\dagger}$ are defined as
\begin{eqnarray}
\hat{\Phi} &=& {1\over{N}} {\rm Tr} L ,~~~~
\hat{\Phi}^{\dagger}  = {1\over{N}} {\rm Tr}L^\dag ,
\end{eqnarray}
with
\begin{eqnarray}
L({\bf x})  &=& {\cal P} \exp\Bigl[
                {i\int^\beta_0 d \tau A_4({\bf x},\tau)}\Bigr],
\end{eqnarray}
where ${\cal P}$ is the path ordering and $A_4 = i A^0 $. 
These are not invariant under the extended ${\mathbb Z}_3$ transformation 
\eqref{ez3}, 
so that their vacuum expectation values do not have the RW periodicity. 
We then introduce the modified Polyakov loop and 
its Hermitian conjugate,
\begin{eqnarray}
\hat{\Psi}_{f}=\exp(i\theta_{f})\hat{\Phi}, \quad 
\hat{\Psi}_{f}^{\dagger}
=\exp(-i\theta_{f})\hat{\Phi}^{\dagger}
\end{eqnarray}
for $f=u, d$. These are invariant under the transformation (\ref{ez3}). 
Their vacuum expectation values $\Psi_{f}=\langle \hat{\Psi}_{f} \rangle$ and 
$\Psi_{f}^{*}=\langle \hat{\Psi}_{f}^{\dagger} \rangle$ 
have the same symmetries as $Z$ 
in \eqref{RW-periodicity}-\eqref{pi-periodicity}; 
note that $\Psi_{f}^{*}$ is the complex conjugate of $\Psi_{f}$ because 
$Z$ is real.

In the chiral limit, QCD has the chiral 
$SU_{\rm L}(2) \times SU_{\rm R}(2)$ symmetry 
when $\mu_{\rm iso} = 0$. 
However, at $\mu_{\rm iso} \neq 0$ this symmetry 
is reduced to $U_{\rm I_3L}(1) \times U_{\rm I_3R}(1)$, where $I_3=\tau_3/2$ 
is the third component of the isospin operator.
Evidently, this symmetry can also be presented as 
$U_{\rm I_3}(1) \times U_{\rm AI_3}(1)$, where $U_{\rm I_3}(1)$ is 
the isospin subgroup and $U_{\rm AI_3}(1)$ is the axial isospin subgroup. 
Quarks are transformed under these subgroups 
as $q \to \exp(i \a \tau_3)q$ and $q \to \exp(i \a \gamma_5 \tau_3)q$,
respectively. 
In the case of $m_u=m_d > 0$, 
only the $U_{\rm I_3}(1)$ symmetry survives.

When QCD vacuum keeps the $U_{\rm v}(1)$ and $U_{\rm I_3}(1)$ symmetries, 
the baryon charge ${\bar B}=V \langle \bar{q}\gamma_4q \rangle$ 
is either zero or integer and 
the isospin charge ${\bar I_3}=V \langle \bar{q}\gamma_4 I_3 q \rangle$ 
is also either zero or half-integer. In the partition function $Z$ 
of \eqref{eq:EQ1}, the baryon- and the isospin-charge operator, 
$\bar{q}\gamma_4 q$ and $\bar{q}\gamma_4 I_3 q$, appear through 
the form 
$\exp(2i \theta_{\rm I} \bar{q}\gamma_4 I_3 q + 
i\theta_{\rm q} \bar{q}\gamma_4 q)$. 
Therefore, $\theta_{\rm I}$ and $\theta_{\rm q}$ have periodicities 
\eqref{2pi-periodicity} and \eqref{pi-periodicity}. 
Meanwhile, if the pion condensation occurs, the $U_{\rm I_3}(1)$ symmetry 
is spontaneously broken and hence the isospin charge  
is neither zero nor half-integer anymore. 
In this situation, QCD vacuum does not have 
periodicities \eqref{2pi-periodicity} and \eqref{pi-periodicity}. 
We will then prove that the pion condensation does not take place 
at imaginary $\mu_{\rm iso}$. Son and Stephanov~\cite{Son-Stephanov} 
show for real $\mu_{\rm iso}$ that the pion condensation emerges 
when $|\mu_{\rm iso}|> m_{\pi}$,  where $m_{\pi}$ is the pion 
mass. For simplicity, we take $\mu_{\rm q}=0$, 
because the quark-number chemical potential 
does not break the $U_{\rm I_3}(1)$ symmetry. 
Following their discussion in Ref.~\cite{Son-Stephanov}, 
we use the chiral perturbation theory that is applicable at 
$\mu_{\rm iso}$ smaller than the chiral scale (the $\rho$ meson mass). 
The chiral Lagrangian for pion field $\Sigma \in {\rm SU(2)}$ with 
finite $\mu_{\rm iso}$ is~\cite{Son-Stephanov}
\begin{eqnarray}
{\cal L}_{\rm eff}=\frac{f_\pi^2}{4}{\rm Tr}
\nabla_\nu \Sigma \nabla_\nu \Sigma^{\dagger}
-\frac{m_\pi^2 f_\pi^2}{2}{\rm Re Tr}\Sigma
\label{eff-L}
\end{eqnarray}
with flavor covariant derivatives 
\begin{eqnarray}
&\nabla_0 \Sigma=\partial_0 \Sigma - \frac{\mu_{\rm iso}}{2} 
\left( \tau_3 \Sigma - \Sigma\tau_3 \right) , 
\notag \\
&\nabla_0 \Sigma^{\dagger}=\partial_0 \Sigma^{\dagger} + 
\frac{\mu_{\rm iso}}{2} 
\left( \Sigma^{\dagger} \tau_3 - \tau_3 \Sigma^{\dagger}   \right) ,  
\end{eqnarray}
where 
$f_\pi$ is the pion decay constant. 
In the effective theory, the condensate ${\bar \Sigma}$ is described by 
\begin{eqnarray}
\bar \Sigma= \tau_0 \cos \a + i \tau_1 \sin \a. 
\label{pi-con-0}
\end{eqnarray}
The tilt angle $\a$ is determined by minimizing 
the vacuum energy (the static part of ${\cal L}_{\rm eff}$) 
\begin{eqnarray}
{\cal L}_{\rm eff}^{\rm st}
=\frac{(f_\pi\mu_{\rm iso})^2}{2}\left[(x-a)^2-1-a^2\right] 
\label{static-L}
\end{eqnarray}
with $x=\cos \a$ and $a=(m_\pi/\mu_{\rm iso})^2$. 
Here, the static part has been obtained by 
inserting \eqref{pi-con-0} into \eqref{eff-L}. 
Noting that $-1 \le x \le 1$, one can find for real $\mu_{\rm iso}$ that 
the static Lagrangian becomes minimum at $x=1$ ($\a=0$) 
when $a > 1$ ($\mu_{\rm iso}<m_\pi$) and  at $x=a$ 
($\a=\arccos(m_{\pi}/\mu_{\rm iso})^2$) 
when $a < 1$ ($\mu_{\rm iso} > m_\pi$)~\cite{Son-Stephanov}. 
The fact that $x=1$ and then ${\bar \Sigma}=\tau_0$ at $\mu_{\rm iso}<m_\pi$ 
means that 
the pion condensation does not take place there.

As expected from \eqref{even-I}, 
the static Lagrangian is $\mu_{\rm iso}$-even and then 
a function of $\mu_{\rm iso}^2$. Hence, the static Lagrangian with 
imaginary isospin chemical potential $\mu_{\rm iso}=i\nu$ is 
given by substituting $i\nu$ for $\mu_{\rm iso}$
in \eqref{static-L}: 
\begin{eqnarray}
{\cal L}_{\rm eff}^{\rm st}=-\frac{(f_\pi\nu)^2}{2}
\left[(x+b)^2-1-b^2\right] 
\label{static-L-I}
\end{eqnarray}
with $b=(m_\pi/\nu)^2$. 
This static Lagrangian is minimum at $x=1$ for any value of $\nu$. 
Therefore, the pion condensation does not occur at 
imaginary $\mu_{\rm iso}$. 
The PNJL model can reproduce this property, as shown in Sec.~\ref{PNJL}. 

The absence of the pion condensation at imaginary $\mu_{\rm iso}$ can be 
understood intuitively as follows. 
For real $\mu_{\rm iso}$, the Bose-Einstein distribution function has an 
infrared divergence at $\mu_{\rm iso}\ge m_{\pi}$. 
This induces the Bose-Einstein Condensation, that is, the pion condensation. 
For imaginary $\mu_{\rm iso}$, such a divergence never happens and then 
no pion condensation occurs.

Putting $x=1$ in (\ref{static-L-I}), one can obtain
\begin{eqnarray}
{\cal L}_{\rm eff}^{\rm st}=-(f_\pi m_\pi)^2 . 
\label{static-L-I-2}
\end{eqnarray}
Thus, in the limit $T\to 0$, 
the static potential (the thermodynamic potential) 
is independent of imaginary $\mu_{\rm iso}$. 
The PNJL model can reproduce this property, as shown later.

\section{PNJL model}
\label{PNJL}

The two-flavor PNJL Lagrangian in Euclidean spacetime is 
\begin{eqnarray}
 {\cal L} = {\bar q}(\gamma_\nu D_\nu - \gamma_4{\hat \mu} + 
              {\hat m}_0 )q 
          + G_{\rm s}[({\bar q}q)^2 
                          +({\bar q}i\gamma_5 {\vec \tau}q)^2] 
              - {\cal U}(\Phi [A],{\Phi} [A]^*,T),~~~
\label{eq:E1}
\end{eqnarray}
where $D_\nu=\partial_\nu-iA_\nu$. The field $A^\nu$ is defined as 
$A^\nu=gA^{a}_{4}{\lambda^a\over{2}}\delta_{\nu4}$
with the gauge field $A^\nu_a$, 
the Gell-Mann matrix $\lambda_a$ and the gauge coupling $g$. 
In the NJL sector, 
$G_{\rm s}$ denotes the coupling constant of the scalar-type 
four-quark interaction. 
The Polyakov potential ${\cal U}$, defined in (\ref{eq:E13}), 
is a function of the Polyakov loop $\Phi$ and its complex conjugate $\Phi^*$. 
In the case of $m_0=\mu_{\rm I}=0$, 
the PNJL Lagrangian has the $SU_{\rm L}(2) \times SU_{\rm R}(2)
\times U_{\rm v}(1) \times SU_{\rm c}(3)$  symmetry. 
In the case of $m_0 \neq 0$ and $\mu_{\rm I} \neq 0$, 
it is reduced to $U_{\rm I_3}(1) \times U_{\rm v}(1) \times SU_{\rm c}(3)$.

In the Polyakov gauge, $L$ can be written in a diagonal form 
in color space~\cite{Fukushima}: 
\begin{eqnarray}
L =  e^{i \beta (\phi_3 \lambda_3 + \phi_8 \lambda_8)}
= {\rm diag} (e^{i \beta \phi_a},e^{i \beta \phi_b},
e^{i \beta \phi_c} ),
\label{eq:E6}
\end{eqnarray}
where $\phi_a=\phi_3+\phi_8/\sqrt{3}$, $\phi_b=-\phi_3+\phi_8/\sqrt{3}$
and $\phi_c=-(\phi_a+\phi_b)=-2\phi_8/\sqrt{3}$. 
The Polyakov loop $\Phi$ is an exact order parameter of the spontaneous 
${\mathbb Z}_3$ symmetry breaking in the pure gauge theory.
Although the ${\mathbb Z}_3$ symmetry is not exact 
in the system with dynamical quarks, it still seems to be a good indicator of 
the deconfinement phase transition. 
Therefore, we use $\Phi$ to define the deconfinement phase transition.

The spontaneous breakings of the chiral and the $U_{\rm I_3}(1)$ symmetry are 
described 
by the chiral condensate $\sigma = \langle \bar{q}q \rangle$ and the charged 
pion condensate~\cite{Zhang}
\begin{eqnarray}
\pi^{\pm}=\frac{\bar {\pi}}{\sqrt{2}}e^{\pm i \varphi}
=\langle \bar{q}i \gamma_5 \tau_{\pm}q \rangle, 
\label{charged-pion}
\end{eqnarray}
where $\tau_{\pm}=(\tau_1\pm i\tau_2)/\sqrt{2}$. 
Since the phase $\varphi$ represents the direction 
of the $U_{\rm I_3}(1)$ symmetry breaking, 
we take $\varphi=0$ for convenience. The pion 
condensate is then expressed by 
\begin{eqnarray}
\bar {\pi}=\langle \bar{q}i \gamma_5 \tau_{1}q \rangle.
\label{pion}
\end{eqnarray}
The mean field (MF) Lagrangian is obtained by~\cite{Zhang} 
\begin{eqnarray}
 {\cal L}_{\rm MF}  & = & {\bar q}(\gamma_\nu D_\nu - \gamma_4{\hat \mu} + 
              M\tau_0 + N i \gamma_5 \tau_{1})q \notag\\
            &\hspace{3mm}&  - G_{\rm s}[\sigma^2 +{\bar \pi}^2] 
              - {\cal U} 
             \label{MF-L}
\end{eqnarray}
where $M=m_0-2 G_{\rm s} \sigma$ and 
$N=-2 G_{\rm s} {\bar \pi}$. 
Performing the path integral in the PNJL partition function 
\begin{eqnarray}
Z_{\rm PNJL}=\int Dq D\bar{q}
\exp\left[ - \int d^4 x {\cal L}_{\rm MF} \right] , 
\label{PNJL-Z}
\end{eqnarray}
one can obtain the thermodynamic potential $\Omega$ 
(per unit volume), 
\begin{eqnarray}
\Omega &=&-T\ln(Z_{\rm PNJL})/V \notag\\
&=& -2\sum_{i=\pm}\int \frac{d^3{\rm p}}{(2\pi)^3}
\Bigl[ 3 E_{i}({\rm p}) 
+ \frac{1}{\beta}\ln~ [1+3(\Phi+\Phi^{*}e^{-\beta E_{i}^-({\bf p})}) 
e^{-\beta E_{i}^-({\bf p})}+ e^{-3\beta E_{i}^- ({\bf p})}]
\notag\\
&+& \frac{1}{\beta}\ln~ [1+3(\Phi^{*}+{\Phi e^{-\beta E_{i}^+({\bf p})}}) 
e^{-\beta E_{i}^+({\bf p})}+ e^{-3\beta E_{i}^+({\bf p})}]\Bigl]
+ G_{\rm s}[\sigma^2 +{\bar \pi}^2]+{\cal U} 
\label{eq:E12-pi} 
\end{eqnarray}
with
\begin{eqnarray}
E_{\pm}({\rm p})
=\sqrt{(E({\rm p})\pm\mu_{\rm I})^2+N^2}, 
\end{eqnarray}
$E_{\pm}^\pm({\rm p})=E_{\pm}({\rm p})\pm \mu_{\rm q}$ 
and $E({\rm p})=\sqrt{{\bf p}^2+M^2}$. Obviously, $\Omega$ does not have
discrete symmetries \eqref{2pi-periodicity} and \eqref{pi-periodicity}, 
when ${\bar \pi} \neq 0$.

On the right-hand side of \eqref{eq:E12-pi}, only the first term diverges, and 
it is then regularized by the three-dimensional momentum 
cutoff $\Lambda$~\cite{Fukushima,Ratti}. 
The parameter set, $\Lambda =631.5$~MeV, 
$G_{\rm s}=5.498$~[GeV$^{-2}]$ and $m_0=5.5$~MeV, 
can reproduce the pion decay constant $f_{\pi}=93.3$~MeV and 
the pion mass $M_{\pi}=138$~MeV at $T=0$~\cite{Kashiwa1}. 
We then adopt these values for $\Lambda$, $G_{\rm s}$ and $m_0$. 
We use ${\cal U}$ of Ref.~\cite{Rossner} that is fitted to LQCD data 
in the pure gauge theory at finite $T$~\cite{Boyd,Kaczmarek}: 
\begin{eqnarray}
&&{\cal U} = T^4 \Bigl[-\frac{a(T)}{2} {\Phi}^*\Phi
+ b(T)\ln(1 - 6{\Phi\Phi^*}  + 4(\Phi^3+{\Phi^*}^3)
            - 3(\Phi\Phi^*)^2 )\Bigr],~~~\label{eq:E13}\\
&&~~~~~~a(T) = a_0 + a_1\Bigl(\frac{T_0}{T}\Bigr)
                 + a_2\Bigl(\frac{T_0}{T}\Bigr)^2,
 ~~b(T)=b_3\Bigl(\frac{T_0}{T}\Bigr)^3~~~\label{eq:E14}
\end{eqnarray}
where parameters are summarized in Table I.  
The Polyakov potential yields a first-order deconfinement phase transition at 
$T=T_0$ in the pure gauge theory.
The original value of $T_0$ is $270$ MeV determined from the pure gauge 
LQCD data, but the PNJL model with this value of $T_0$ yields somewhat 
larger value of the pseudocritical temperature at zero chemical potential than 
the full LQCD simulation~\cite{Karsch,Kaczmarek2} predicts. 
Therefore, we rescale $T_0$ to 212~MeV~\cite{Sakai2}. 

\begin{table}[h]
\begin{center}
\begin{tabular}{llllll}
\hline
~~~~~$a_0$~~~~~&~~~~~$a_1$~~~~~&~~~~~$a_2$~~~~~&~~~~~$b_3$~~~~~
\\
\hline
~~~~3.51 &~~~~-2.47 &~~~~15.2 &~~~~-1.75\\
\hline
\end{tabular}
\caption{
Summary of the parameter set in the Polyakov-potential sector 
determined in Ref.~\cite{Rossner}. 
All parameters are dimensionless. 
}
\end{center}
\end{table}

The classical variables $X=\Phi$, ${\Phi}^*$, $\sigma$ and $\bar {\pi}$ 
satisfy the stationary conditions, 
\begin{eqnarray}
\partial \Omega/\partial X=0. 
\label{eq:SC}
\end{eqnarray}
The solutions of the stationary conditions do not give 
the global minimum of $\Omega$ 
necessarily. There is a possibility 
that they yield a local minimum or even 
a maximum. We then have checked that the solutions yield 
the global minimum when the solutions $X(\theta_{\rm q},\theta_{\rm I})$ 
are inserted into (\ref{eq:E12-pi}).

\begin{figure}[htbp]
\begin{center}
 \includegraphics[width=0.40\textwidth]{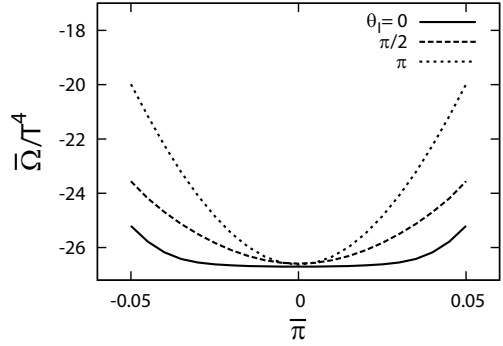}
\end{center}
\caption{Potential surface as a function of ${\bar \pi}$ at 
$T=175$~MeV and $\theta_{\rm q}=0$. 
The solid, dashed and dotted curves denote 
three cases of $\theta_{\rm I}=0, \pi/2$ and $\pi$, respectively. 
}
\label{no-pi}
\end{figure}

Now we numerically confirm that the pion condensation does not occur
at imaginary $\mu_{\rm I}$. For simplicity, we set $\mu_{\rm q}=0$, 
since the quark-number chemical potential does not break the $U_{\rm I_3}(1)$ 
symmetry. 
For this purpose, we search for the potential minimum by varying 
$\Phi$, ${\Phi}^*$ and $\sigma$ with $\bar {\pi}$ fixed. 
The potential surface ${\bar \Omega}({\bar \pi})$ thus obtained 
is a function of $\bar {\pi}$ and drawn 
in Fig.~\ref{no-pi}, where $T$ is taken to be $175$~MeV. 
Three cases of $\theta_{\rm I}=0, \pi/2$ and $\pi$ are represented by 
the solid, dashed and dotted curves, respectively. 
For the three cases, the global minimum is always 
located at $\bar {\pi}=0$. 
The curvature around the minimum becomes large 
as $\theta_{\rm I}$ increases. This means that the vacuum 
becomes more stable for larger $\theta_{\rm I}$.

Therefore, we can set ${\bar \pi}=0$. In this situation, 
the transformation \eqref{1st-trans} reduces 
${\cal L}_{\rm MF}$ of \eqref{MF-L} to 
\begin{eqnarray}
 {\cal L}_{\rm MF}  = {\bar q}(\gamma_\nu D_\nu + M \tau_0)q 
             - G_{\rm s}\sigma^2 - {\cal U} 
\label{MF-L-2}
\end{eqnarray}
with the boundary conditions \eqref{BC1}. 
Note that this procedure breaks down if ${\bar \pi} \neq 0$, since 
the operator $\bar{q}i \gamma_5 \tau_{1}q$ is not invariant under 
the transformation \eqref{1st-trans}. 
Following Sec.~\ref{Symmetries},  
one can show that the thermodynamic potential $\Omega$ has 
the same symmetries as $Z$ in \eqref{RW-periodicity}-\eqref{pi/3}. 
This statement is proven below more explicitly.

Under the fact that ${\bar \pi}=0$, 
$\Omega$ is reduced to a simpler form 
\begin{eqnarray}
\Omega &=& -2\sum_{f=u,d}\int \frac{d^3{\rm p}}{(2\pi)^3}
         \Bigl[ 3 E({\rm p}) 
        + \frac{1}{\beta}
         \ln~ [1 + 3(\Phi+\Phi^{*} e^{-\beta E_{f}^-({\bf p})}) 
        e^{-\beta E_{f}^-({\bf p})}+ e^{-3\beta E_{f}^- ({\bf p})}]
         \nonumber\\
       &+& \frac{1}{\beta} 
           \ln~ [1 + 3(\Phi^{*}+{\Phi e^{-\beta E_{f}^+({\bf p})}}) 
            e^{-\beta E_{f}^+({\bf p})}+ e^{-3\beta E_{f}^+({\bf p})}]
           \Bigl]
        + G_{\rm s} \sigma^2+{\cal U}. 
\label{eq:E12} 
\end{eqnarray}
where $E_{f}^\pm({\rm p})=E({\rm p})\pm \mu_{f}
=E({\rm p})\pm i\theta_{f}/\beta$. 
Obviously, $\Omega$ has 
discrete symmetries \eqref{2pi-periodicity} and \eqref{pi-periodicity}. 
In the limit of $T=0$, on 
the right-hand side of \eqref{eq:E12} the first term 
including $3 E({\rm p})$ and the term 
$G_{\rm s} \sigma^2+{\cal U}$ survive, and hence $\Omega$ has no
$\mu_{\rm q}$ and $\mu_{\rm I}$ dependences there.

The thermodynamic potential $\Omega$ of \eqref{eq:E12} is not 
invariant under the ${\mathbb Z}_3$ transformation, 
\begin{eqnarray}
\Phi \to \Phi e^{-i{2\pi k/{3}}} \;,\quad
\Phi^{*} \to \Phi^{*} e^{i{2\pi k/{3}}} \;, 
\end{eqnarray}
although 
${\cal U}$ of (\ref{eq:E13}) is invariant. 
Instead of the ${\mathbb Z}_3$ symmetry, however, 
$\Omega$ is invariant under the extended ${\mathbb Z}_3$ transformation, 
\begin{eqnarray}
&e^{\pm i \theta_{q}} \to e^{\pm i \theta_{\rm q}} e^{\pm i{2\pi k\over{3}}},
\quad 
\Phi \to \Phi e^{-i{2\pi k\over{3}}}, 
\quad 
&\Phi^{*} \to \Phi^{*} e^{i{2\pi k\over{3}}} .
\label{eq:K2}
\end{eqnarray}
This is easily understood as follows. 
It is convenient to introduce the modified Polyakov loop 
$\Psi_{f} \equiv e^{i\theta_{f}}\Phi$ and 
$\Psi_{f}^{*} \equiv e^{-i\theta_{f}}\Phi^{*}$ 
that are invariant under the transformation (\ref{eq:K2}) 
and have the same symmetries as $Z$ 
in \eqref{2pi-periodicity}-\eqref{pi-periodicity}. 
The extended ${\mathbb Z}_3$ transformation is then 
rewritten into 
\begin{eqnarray}
e^{\pm i \theta_{\rm q}} \to e^{\pm i \theta_{\rm q}}
 e^{\pm i{2\pi k\over{3}}}, \quad
\Psi_{f} \to \Psi_{f}, \quad
\Psi_{f}^{*} \to \Psi_{f}^{*} ,
\label{eq:K2'}
\end{eqnarray}
and $\Omega$ is also into  
\begin{eqnarray}
\Omega &=& -2 \sum_{f=u,d} \int \frac{d^3{\rm p}}{(2\pi)^3}
          \Bigl[ 3 E ({\rm p}) 
          + \frac{1}{\beta}\ln~ [1 + 3\Psi_{f} e^{-\beta E({\bf p})}
        + 3\Psi_{f}^{*}e^{-2\beta E({\bf p})}e^{3i\theta_{f}}
          + e^{-3\beta E({\bf p})}e^{3i\theta_{f}}] \notag\\
       &+& \frac{1}{\beta} 
           \ln~ [1 + 3\Psi_{f}^{*} e^{-\beta E({\bf p})}
          + 3\Psi_{f} e^{-2\beta E({\bf p})}e^{-3i\theta_{f}}
        + e^{-3\beta E({\bf p})}e^{-3i\theta_{f}}]
	      \Bigl]+G_{\rm s} \sigma^2+ {\cal U} .
\label{eq:K3} 
\end{eqnarray}
Obviously, $\Omega$ is invariant under the extended ${\mathbb Z}_3$ 
transformation \eqref{eq:K2'}, since 
it is a function of only extended ${\mathbb Z}_3$ invariant 
quantities, 
$e^{3i\theta_{f}}=e^{3i\theta_{\rm q}}e^{\pm 3i\theta_{\rm I}}$ 
($+$ for u-quark and $-$ for d-quark) 
and ${X}(=\Psi_{f}, \Psi_{f}^{*},\sigma$). 
The explicit $\theta_{\rm q}$ dependence appears only 
through the factor $e^{3i\theta_{\rm q}}$ in (\ref{eq:K3}). 
Hence, the stationary conditions (\ref{eq:SC}) show that 
${X}={X}(e^{3i\theta_{\rm q}})$. 
Inserting the solutions back to (\ref{eq:K3}), one can see that 
$\Omega=\Omega(e^{3i\theta_{\rm q}})$. 
Thus, ${X}$ and $\Omega$ have the 
RW periodicity, 
\begin{eqnarray}
{X}(\theta_{\rm q}+\frac{2\pi k}{3})={X}(\theta_{\rm q}), 
\quad 
\Omega(\theta_{\rm q} +\frac{2\pi k}{3})=\Omega(\theta_{\rm q}), 
\end{eqnarray}
and then
\begin{eqnarray}
\Phi(\theta_{\rm q}+\frac{2\pi k}{3})&=&e^{-i2\pi k/3}\Phi(\theta_{\rm q}).
\end{eqnarray}

The thermodynamic potential $\Omega$ of \eqref{eq:K3} is invariant under 
the transformation  $\theta_{\rm I} \to -\theta_{\rm I}$, indicating that 
$\Omega$ is $\theta_{\rm I}$-even. The thermodynamic potential $\Omega$ is 
also invariant under the $\theta_{\rm q} \to -\theta_{\rm q}$ transformation, 
if $\Psi_f$ is replaced by $\Psi_f^*$. This means that the solutions of the 
stationary condition \eqref{eq:SC} satisfy
\begin{eqnarray}
\Psi_f(\theta_{\rm q})=\Psi_f^*(-\theta_{\rm q}) ,
\label{Psi-relation}
\end{eqnarray}
indicating that $\Omega$ is $\theta_{\rm q}$-even. 
Furthermore, $\Omega$ of \eqref{eq:K3} satisfies the symmetries 
\eqref{2pi-periodicity} and \eqref{pi-periodicity}. 
These properties, together with the RW periodicity, guarantee 
that $\Omega$ of PNJL has the 
same symmetries as $Z$ of QCD in \eqref{RW-periodicity}-\eqref{pi/3}. 
The symmetries \eqref{RW-periodicity}-\eqref{pi/3} are visualized by 
numerical calculations in Sec. \ref{Numerical-results}.

Particularly at $\theta_{\rm I}=\pi/2$, 
$\Omega$ has a periodicity of $\pi/3$ in $\theta_{\rm q}$, because  
taking $\theta_{\rm I}$ to $\pi/2$ in (15) leads to  
\begin{eqnarray}
\Omega(\theta_{\rm q},\pi/2)=\Omega(\theta_{\rm q}+\pi/3,\pi/2).  
\label{pi/3-2}
\end{eqnarray}
As shown in \eqref{eq:E12}, $\Omega$ is a sum of 
the thermodynamic potential $\Omega_u(\theta_u)$ for $u$-quark and 
that $\Omega_d(\theta_d)$ for $d$-quark, i.e., 
$\Omega=\Omega_u(\theta_u)+\Omega_d(\theta_d)$, and $\Omega_f(\theta_f)$ 
is a periodic even function of $3\theta_f$. 
Hence, $\Omega_f$ can be expanded by 
$\cos(3k\theta_f)$ with integer $k$. We then have 
\begin{eqnarray}
\Omega=\sum_k a_k \left[
\cos(3k\theta_u) + \cos(3k\theta_d) \right] . 
\label{k-expansion}
\end{eqnarray}
At lower temperature such as $T \la 2 T_c$, 
where $T_c$ is the pseudocritical temperature 
of the deconfinement transition 
at $\mu_{\rm q}=\mu_{\rm I}=0$, 
the coefficients $\{ a_k \}$ of the expansion have the property 
that the $a_k$ with $k \ge 2$ are 
small~\cite{Kashiwa3}. In particular when $\theta_{\rm I}=\pi/2$, 
$\Omega$ is reduced to 
\begin{eqnarray}
\Omega \approx 
2 a_0 + a_1 \left[
\cos(3\theta_{\rm q}+3\pi/2) + \cos(3\theta_{\rm q}-3\pi/2) \right]
= 2 a_0 
\label{constant}
\end{eqnarray}
for any $\theta_{\rm q}$. 
Accordingly, when $\theta_{\rm I}=\pi/2$, 
$\Omega$ has a periodicity of $\pi/3$ in $\theta_{\rm q}$, but 
the dependence is quite weak. 
This property is also visualized by numerical 
calculations in Sec. \ref{Numerical-results}.


\section{Numerical results}
\label{Numerical-results}

\subsection{$\theta_{\rm q}$ dependence}

$\theta_{\rm q}$ dependence of $\Omega$, 
the quark number density $n_{\rm q}=-d\Omega/d(iT\theta_{\rm q})$ and 
the isospin number density $n_{\rm I}=-d\Omega/d(iT\theta_{\rm I})$ 
is investigated in this subsection. 
The thermodynamic potential $\Omega$ is real 
and $\theta_{\rm q}$-even, so that $n_{\rm q}$ and $n_{\rm I}$ are pure 
imaginary. 
$n_{\rm q}$ is $\theta_{\rm q}$-odd and $\theta_{\rm I}$-even. 
$n_{\rm I}$ is $\theta_{\rm q}$-even and $\theta_{\rm I}$-odd.

As for $\theta_{\rm I}=0$, it is known that, 
at temperature above $T_{\rm RW}=1.1T_c=190$~MeV~\cite{Sakai2}, 
$d\Omega/d\theta_{\rm q}$ is discontinuous at 
$\theta_{\rm q}=\pi/3$ mod $2\pi/3$; note that $T_c=173$~MeV 
in the present PNJL calculation. 
This discontinuity is called the RW phase transition. 
At such higher temperatures, three ${\mathbb Z}_3$ vacua emerge 
alternatively in variation of $\theta_{\rm q}$, that is, 
the first vacuum appears in the region (I) $-\pi/3 < \theta_{\rm q} < \pi/3$, 
the second one in the region (II) $\pi/3 < \theta_{\rm q} < \pi$ and 
the third one in the region (III) $-\pi < \theta_{\rm q} < -\pi/3$. 
As a result of this mechanism, 
$d\Omega/d\theta_{\rm q}$ becomes 
discontinuous at boundaries of the three regions~\cite{RW,Kouno}. 
The charge conjugation is an exact symmetry on the boundaries. It is 
preserved below $T_{\rm RW}$, but spontaneously broken 
above $T_{\rm RW}$~\cite{Kouno}.

\begin{figure}[htbp]
\begin{center}
 \includegraphics[width=0.30\textwidth,angle=-90]{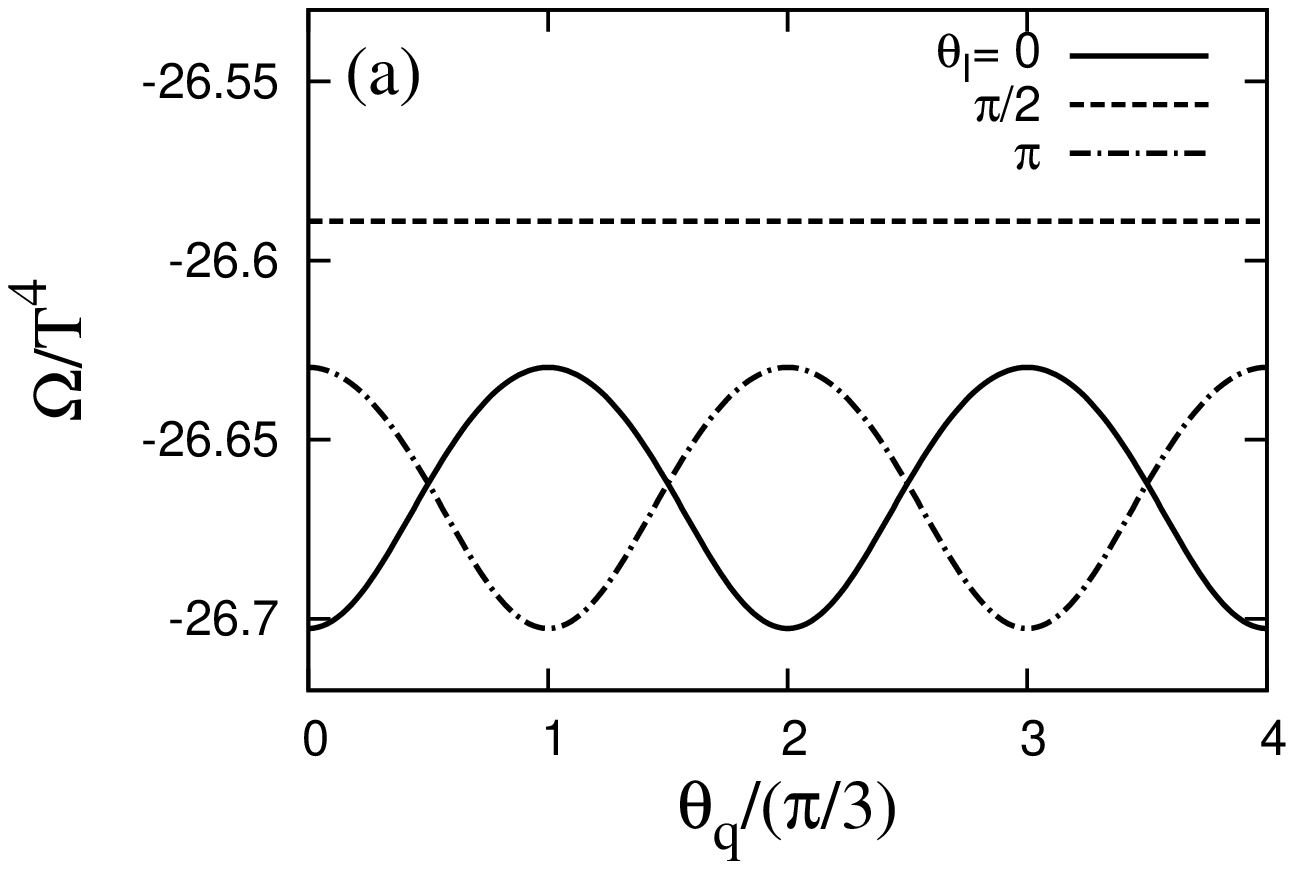}
 \includegraphics[width=0.30\textwidth,angle=-90]{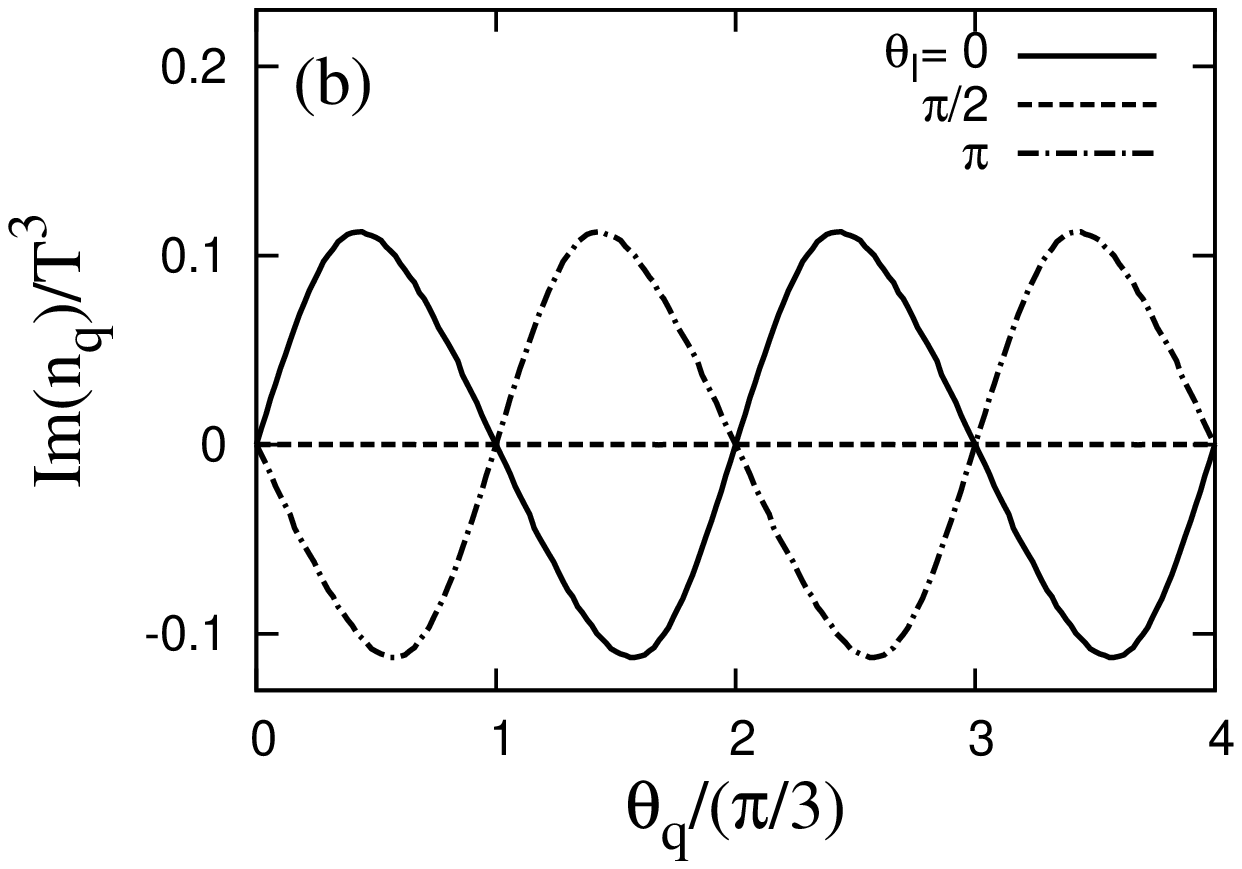}
 \includegraphics[width=0.30\textwidth,angle=-90]{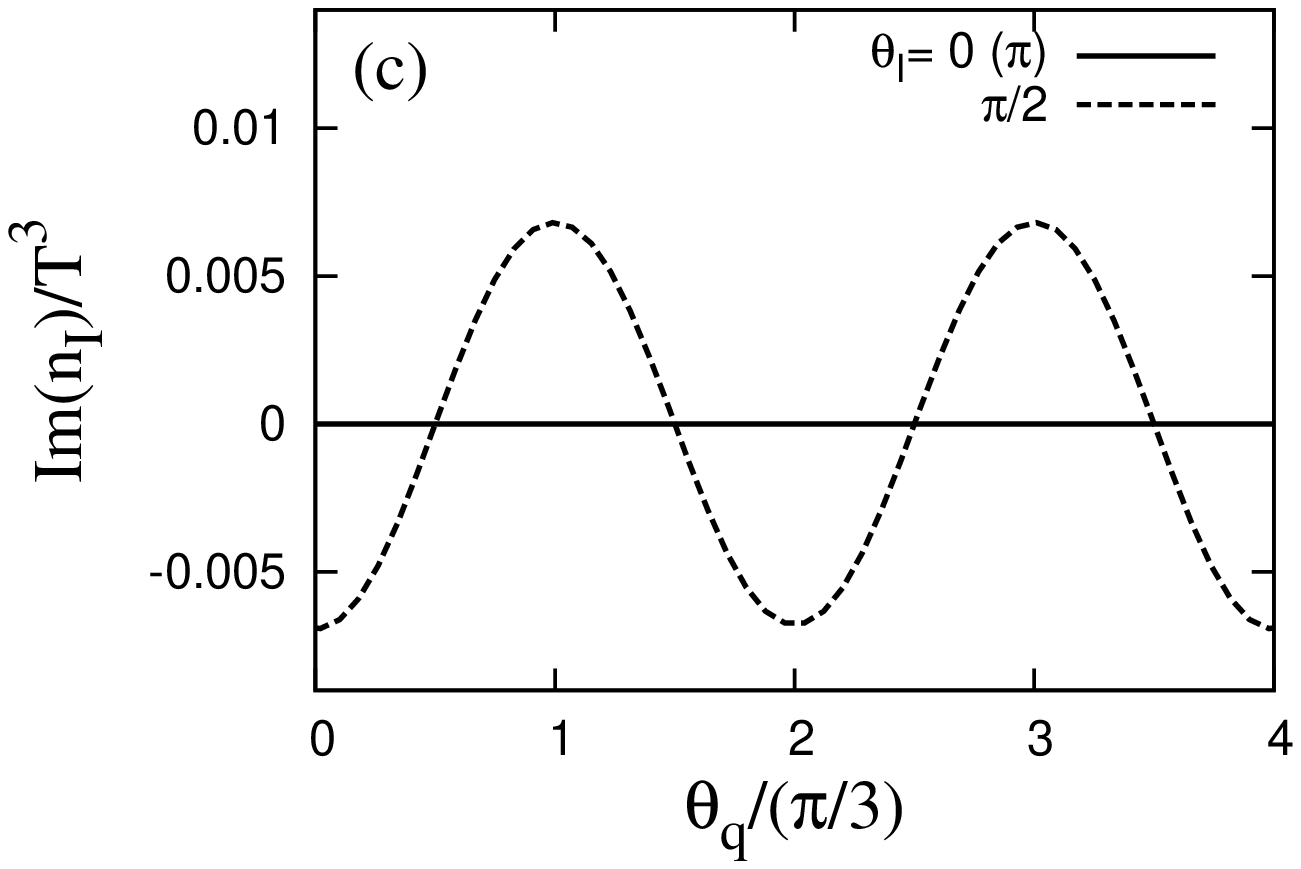}
\end{center}
\caption{$\theta_{\rm q}$ dependence of (a) $\Omega/T^4$, 
(b) ${\rm Im}[n_{\rm q}]/T^3$ and (c) 
${\rm Im}[n_{\rm I}]/T^3$ at $T=175$~MeV. 
Three cases of $\theta_{\rm I}=0, \pi/2$ and $\pi$ are represented by 
solid, dashed and dot-dashed curves, respectively. 
}
\label{175T-Q}
\end{figure}

Now, we consider $T=175$~MeV as a typical temperature below 
$T_{\rm RW}$. 
Figure~\ref{175T-Q} presents $\theta_{\rm q}$ dependence of 
$\Omega/T^4$, the imaginary parts ${\rm Im}[n_{\rm q}/T^3]$ and 
${\rm Im}[n_{\rm I}/T^3]$ for three cases of 
$\theta_{\rm I}=0, \pi/2$ and $\pi$. These quantities have the RW periodicity 
and are smooth at any $\theta_{\rm q}$, 
as expected. Further, $\Omega$ and $n_{\rm I}$ are $\theta_{\rm q}$-even, 
while $n_{\rm q}$ is $\theta_{\rm q}$-odd.  
In the case of $\theta_{\rm I}=\pi/2$, $\Omega$ is almost constant and 
${\rm Im}[n_{\rm q}]$ is then nearly zero, as expected from \eqref{constant}; 
precisely, they have a periodicity of $\pi/3$, but the $\theta_{\rm q}$ 
dependence is quite weak. 
Meanwhile, ${\rm Im}[n_{\rm I}]$ is zero when $\theta_{\rm I}=0$ and $\pi$, 
because it is $\theta_{\rm I}$-odd and satisfies \eqref{pi-periodicity}. 
As for the case of $\theta_{\rm I}=\pi/2$, 
${\rm Im}[n_{\rm I}]$ has the RW periodicity clearly.

\begin{figure}[htbp]
\begin{center}
 \includegraphics[width=0.40\textwidth]{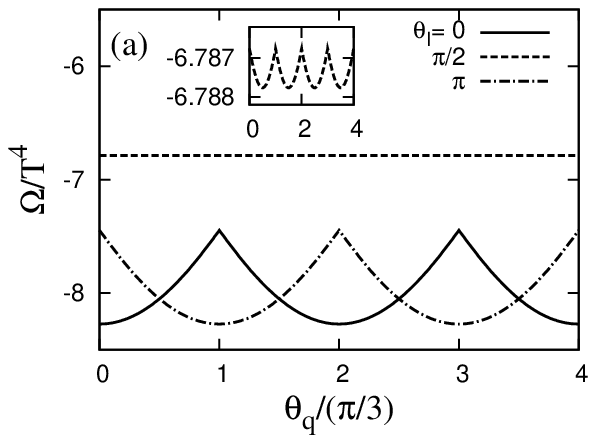}
 \includegraphics[width=0.40\textwidth]{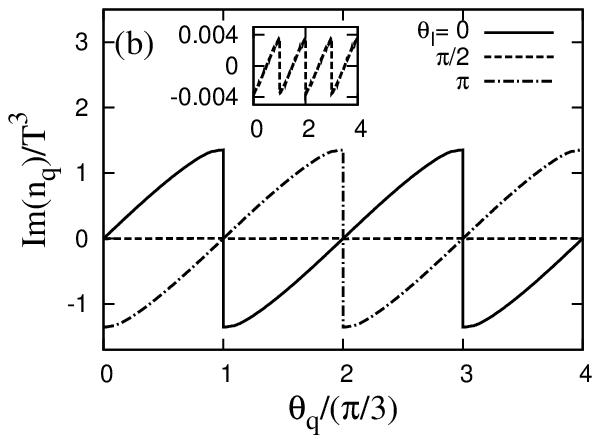}
 \includegraphics[width=0.40\textwidth]{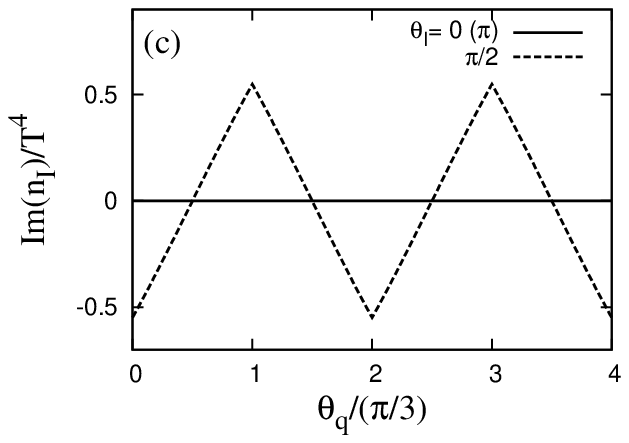}
\end{center}
\caption{
$\theta_{\rm q}$ dependence of (a) $\Omega/T^4$, (b)
${\rm Im}[n_{\rm q}]/T^3$ and 
(c) ${\rm Im}[n_{\rm I}]/T^3$ at $T=250$~MeV. 
Three cases of $\theta_{\rm I}=0, \pi/2$ and $\pi$ are taken. 
In the panel (c), the solid and dot-dashed lines agree with 
the $x$ axis. 
Definitions of curves are the same as in Fig.~\ref{175T-Q}. 
In the insets, these quantities at $\theta_{\rm I}=\pi/2$ are 
magnified. 
}
\label{250T-Q}
\end{figure}

Figure \ref{250T-Q} shows the same quantities as Fig.~\ref{175T-Q}, but 
its temperature is $T=250$MeV higher than $T_{\rm RW}$. 
The RW periodicity is seen also in this figure. 
In the case of $\theta_{\rm I}=0$, $\Omega$ and $n_{\rm I}$
have cusps at $\theta_{\rm q}=\pi/3$ mod $2\pi/3$, 
while $n_{\rm q}$ is discontinuous there. 
This discontinuity means the RW phase transition. 
When $\theta_{\rm I}=\pi/2$, $\Omega$ is almost constant, 
as expected from \eqref{constant}, and 
${\rm Im}[n_{\rm q}]$ is tiny everywhere. In the insets where  
$\Omega$ and ${\rm Im}[n_{\rm q}]$ at $\theta_{\rm I}=\pi/2$ are 
magnified, as expected from \eqref{pi/3}, $\Omega$ and ${\rm Im}[n_{\rm I}]$ 
have cusps at $\theta_{\rm q}=0$ mod $\pi/3$, while 
${\rm Im}[n_{\rm q}]$ is discontinuous there. 
As for the case of $\theta_{\rm I}=\pi/2$, thus, the RW phase transition 
occurs at $\theta_{\rm q}=0$ mod $\pi/3$. 
Equation \eqref{pi/3} yields a relation 
\begin{eqnarray}
\Omega(\theta_{\rm q}-\pi/3,\theta_{\rm I}+\pi)
=\Omega(\theta_{\rm q},\theta_{\rm I}) .
\label{shift}
\end{eqnarray}
As a consequence of this symmetry, in Figs.~\ref{175T-Q} and \ref{250T-Q}, 
the dot-dashed curves are obtained by shifting the corresponding solid curves 
by $\pi/3$ in the $\theta_{\rm q}$ direction.

The discontinuity between the right- and left-hand limits of 
${\rm Im}[n_{\rm q}(\theta_{\rm q})]$ as $\theta_{\rm q}$ approaches 
$\pi/3$, i.e., 
${\rm Im}[n_{\rm q}(+\pi/3)-n_{\rm q}(-\pi/3)]$,  
 decreases as $\theta_{\rm I}$ increases from 0 and 
disappears at $\theta_{\rm I}= \pi/2 + \delta(T)$, 
as shown later in Fig. \ref{250T-I}(b) and \ref{Q-I}(e). 
Here, $\delta(T)$ numerically obtained is a small number depending on $T$ 
weakly:
\begin{eqnarray}
\delta(T) = 0.00016 \times (T-250)
\label{delta}
\end{eqnarray}
for $T \ge 212~{\rm MeV}$. 
Since the discontinuity of ${\rm Im}[n_{\rm q}(\theta_{\rm q})]$ 
means the RW phase transition, 
$\theta_{\rm I}= \pi/2 + \delta(T)$ 
represents a location of an endpoint of the RW phase transition. 
Further discussion on the endpoint is made later in Sec.~\ref{I-T-plane}.

\begin{figure}[htbp]
\begin{minipage}{0.47\textwidth}
 \includegraphics{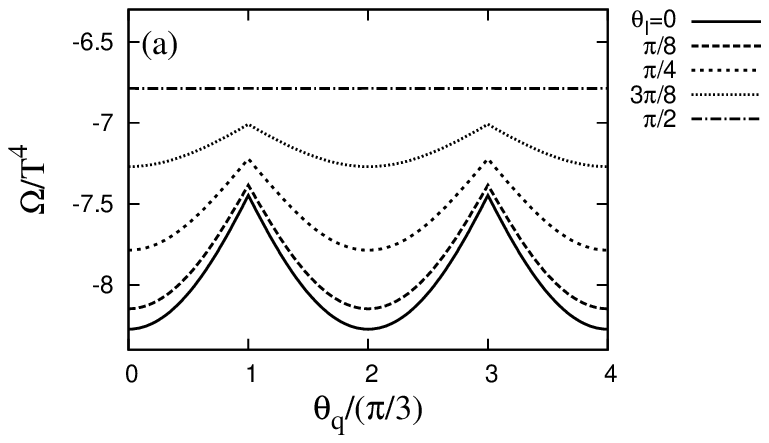}
\end{minipage}
\begin{minipage}{0.47\textwidth}
 \includegraphics{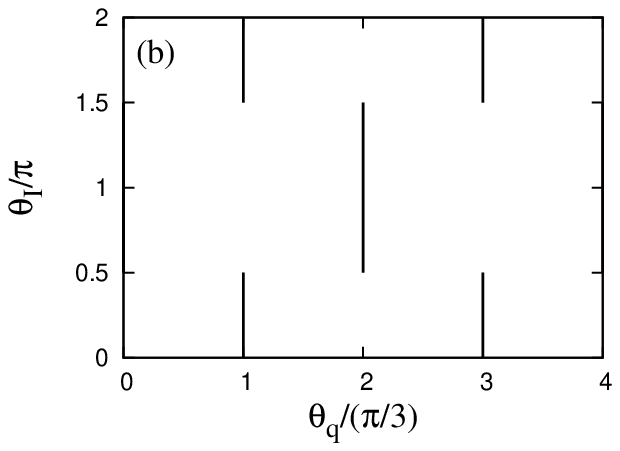}
\end{minipage}
\caption{
(a) $\theta_{\rm q}$ dependence of 
$\Omega/T^4$ for five cases of 
$\theta_{\rm I}=0$ (solid curve), $\pi/8$ (thick dashed curve), 
$\pi/4$ (thin dashed curve), 
$3\pi/8$ (dotted curve) and $\pi/2$ (dot-dashed curve). 
(b) Phase diagram in $\theta_{\rm q}$-$\theta_{\rm I}$ plane. 
The solid lines represent the RW phase transition. 
For both the panels, $T=250$~MeV. 
}
\label{Thetaq-ThetaI-1}
\end{figure}

For simplicity, our discussion begins with the case of $T=250$~MeV, 
since $\delta(T)=0$ there. 
Figure~\ref{Thetaq-ThetaI-1} (a) presents $\theta_{\rm q}$ dependence of 
$\Omega/T^4$ for five cases of 
$\theta_{\rm I}=0$, $\pi/8$, $\pi/4$, $3\pi/8$ and $\pi/2$. 
These results show that 
the RW phase transition occurs at $\theta_{\rm q}=\pi/3$ mod $2\pi/3$ when 
$0\le\theta_{\rm I} < \pi/2$. 
Figure~\ref{Thetaq-ThetaI-1} (b) represents 
the location of the RW phase transition 
in $\theta_{\rm q}$-$\theta_{\rm I}$ plane by solid lines.  
As mentioned above, when $0\le\theta_{\rm I} < \pi/2$, 
the RW phase transition occurs at $\theta_{\rm q}=\pi/3$ mod $2\pi/3$. 
This RW phase transition is also seen at $-\pi/2 < \theta_{\rm I} < 0$, 
because $\Omega$ is $\theta_{\rm I}$-even. 
Furthermore, \eqref{shift} indicates 
that the RW phase transition occurs also 
at $\theta_{\rm q}=0$ mod $2\pi/3$ 
when $\pi/2<\theta_{\rm I}<3\pi/2$. 
This is clearly seen in Fig.~\ref{Thetaq-ThetaI-1}.

\begin{figure}[htbp]
\begin{center}
 \includegraphics[width=0.40\textwidth]{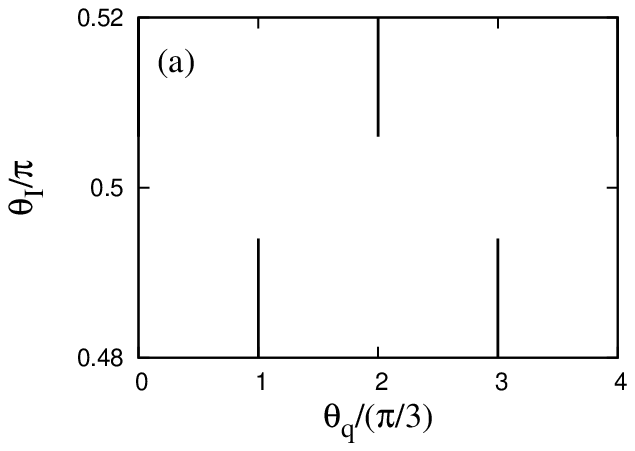}
 \includegraphics[width=0.40\textwidth]{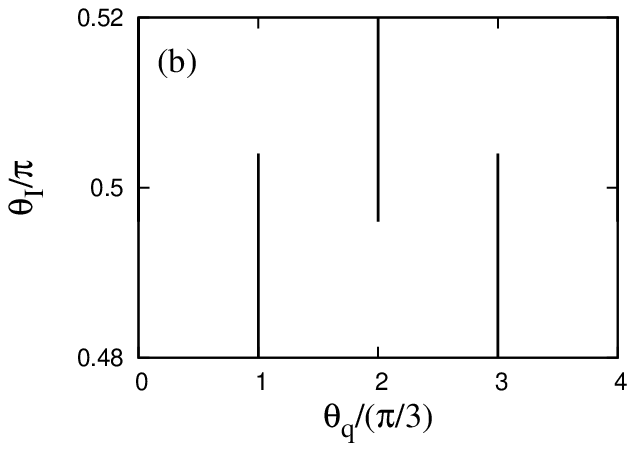}
\end{center}
\caption{Phase diagram in $\theta_{\rm q}$-$\theta_{\rm I}$ plane 
at (a) $T=220$~MeV and (b) $T=300$~MeV in the vicinity of 
$\theta_{\rm I}=\pi/2$. 
The solid lines represent the RW phase transition. 
}
\label{Thetaq-ThetaI-2}
\end{figure}

For other $T$ larger than 212~MeV, $\delta(T)$ is not zero. 
This makes the situation a bit more complicated. 
Following the logic mentioned above, we can 
find that the RW phase transition occurs 
at $\theta_{\rm q}=\pi/3$ mod $2\pi/3$ 
when $-\pi/2 - \delta(T)<\theta_{\rm I}< \pi/2 + \delta(T)$
and also at $\theta_{\rm q}=0$ mod $2\pi/3$ 
when $\pi/2-\delta(T)<\theta_{\rm I}<3\pi/2+\delta(T)$. 
This behavior in the vicinity of 
$\theta_{\rm I}=\pi/2$ is confirmed 
in Fig.~\ref{Thetaq-ThetaI-2} that presents 
the phase diagram in $\theta_{\rm q}$-$\theta_{\rm I}$ plane 
at (a) $T=220$~MeV and (b) $T=300$~MeV. 
Note that $\delta(T)$ is negative in panel (a), but positive in panel (b).

\subsection{$\theta_{\rm I}$ dependence}

$\theta_{\rm I}$ dependence of $\Omega$, 
$n_{\rm q}$ and $n_{\rm I}$ is investigated in this subsection. 
Equations \eqref{2pi-periodicity} and \eqref{even-I} lead to a relation 
\begin{eqnarray}
\Omega(\theta_{\rm q}, \pi-\theta_{\rm I})
=\Omega(\theta_{\rm q}, \theta_{\rm I}-\pi)
=\Omega(\theta_{\rm q}, \theta_{\rm I}+\pi). 
\label{pi-periodicity-2}
\end{eqnarray}
Thus, $\theta_{\rm I}$ dependence of $\Omega$ is 
symmetric with respect to the axis $\theta_{\rm I}=\pi$. 
Differentiating \eqref{pi-periodicity-2} with respect to 
$\theta_{\rm q}$, one can see that $\theta_{\rm q}$-odd quantities such as 
$n_{\rm q}$ have the same symmetry as $\theta_{\rm q}$-even ones such as 
$\Omega$: 
\begin{eqnarray}
n_{\rm q}(\theta_{\rm q}, \pi-\theta_{\rm I})
=n_{\rm q}(\theta_{\rm q}, \theta_{\rm I}-\pi)
=n_{\rm q}(\theta_{\rm q}, \theta_{\rm I}+\pi). 
\label{pi-periodicity-3}
\end{eqnarray}
In contrast, differentiating \eqref{pi-periodicity-2} with respect to 
$\theta_{\rm q}$ leads to the fact that 
the $\theta_{\rm I}$ dependence of the $\theta_{\rm I}$-odd 
quantities such as $n_{\rm I}$ is asymmetric with respect to the axis 
$\theta_{\rm I}=\pi$: 
\begin{eqnarray}
-n_{\rm I}(\theta_{\rm q}, \pi-\theta_{\rm I})
=n_{\rm I}(\theta_{\rm q}, \theta_{\rm I}+\pi). 
\label{pi-periodicity-4}
\end{eqnarray}
Taking $\theta_{\rm q}$ to $\pi/6$ in \eqref{pi/3}, one can find 
\begin{eqnarray}
\Omega(\pi/6, \theta_{\rm I})=\Omega(\pi/6, \theta_{\rm I}+\pi) , 
~~~n_{\rm I}(\pi/6, \theta_{\rm I})=n_{\rm I}(\pi/6, \theta_{\rm I}+\pi) , 
\label{pi/6-1}
\end{eqnarray}
indicating that $\theta_{\rm q}$-even quantities such as $\Omega$ and 
$n_{\rm I}$ have a periodicity of $\pi$ in $\theta_{\rm I}$ when 
$\theta_{\rm q}=\pi/6$. 
Similarly, differentiating \eqref{pi/3} with respect to $\theta_{\rm q}$ and 
setting $\theta_{\rm q}$ to $\pi/6$, we can get 
\begin{eqnarray}
n_{\rm q}(\pi/6, \theta_{\rm I})
=-n_{\rm q}(\pi/6, \theta_{\rm I}+\pi). 
\label{pi/6-2}
\end{eqnarray}
Thus, $n_{\rm q}(\pi/6, \theta_{\rm I})$ has an anti-periodicity 
of $\pi$ in $\theta_{\rm I}$, that is, 
the sign of $n_{\rm q}$ is changed by the transformation 
$\theta_{\rm I} \to \theta_{\rm I}+\pi$. 
These properties of \eqref{pi-periodicity-2}-\eqref{pi/6-2} are seen below 
in Figs.~\ref{175T-I} and \ref{250T-I}.

\begin{figure}[htbp]
\begin{center}
 \includegraphics[width=0.30\textwidth,angle=-90]{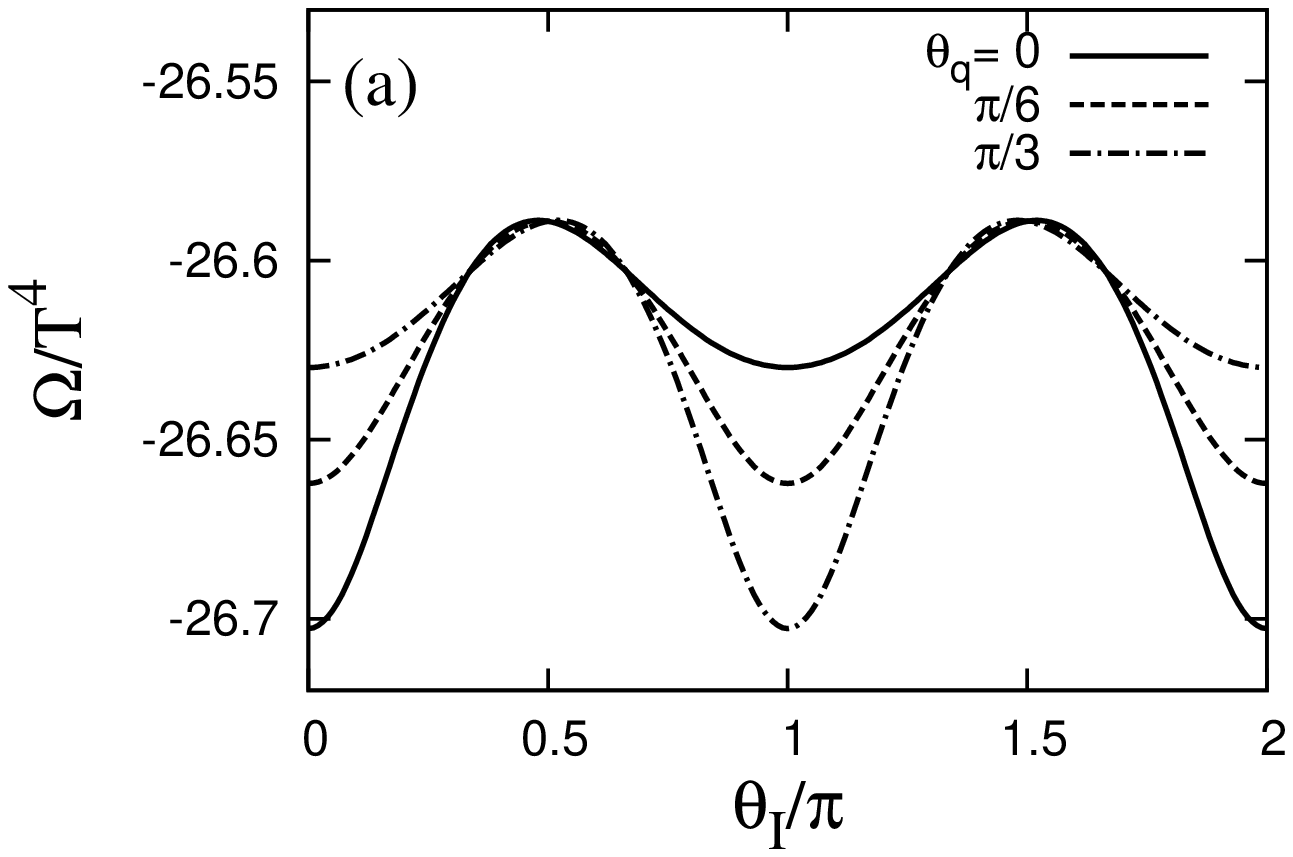}
 \includegraphics[width=0.30\textwidth,angle=-90]{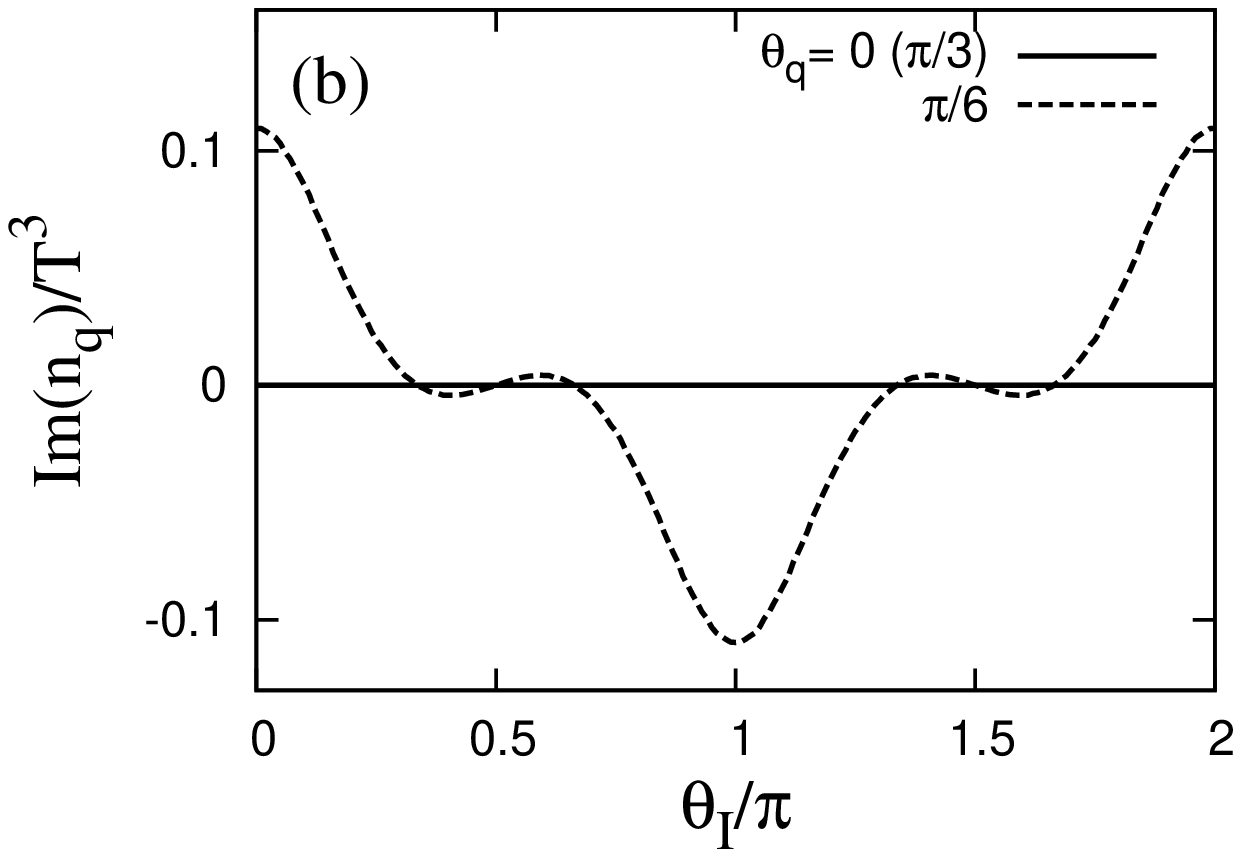}
 \includegraphics[width=0.30\textwidth,angle=-90]{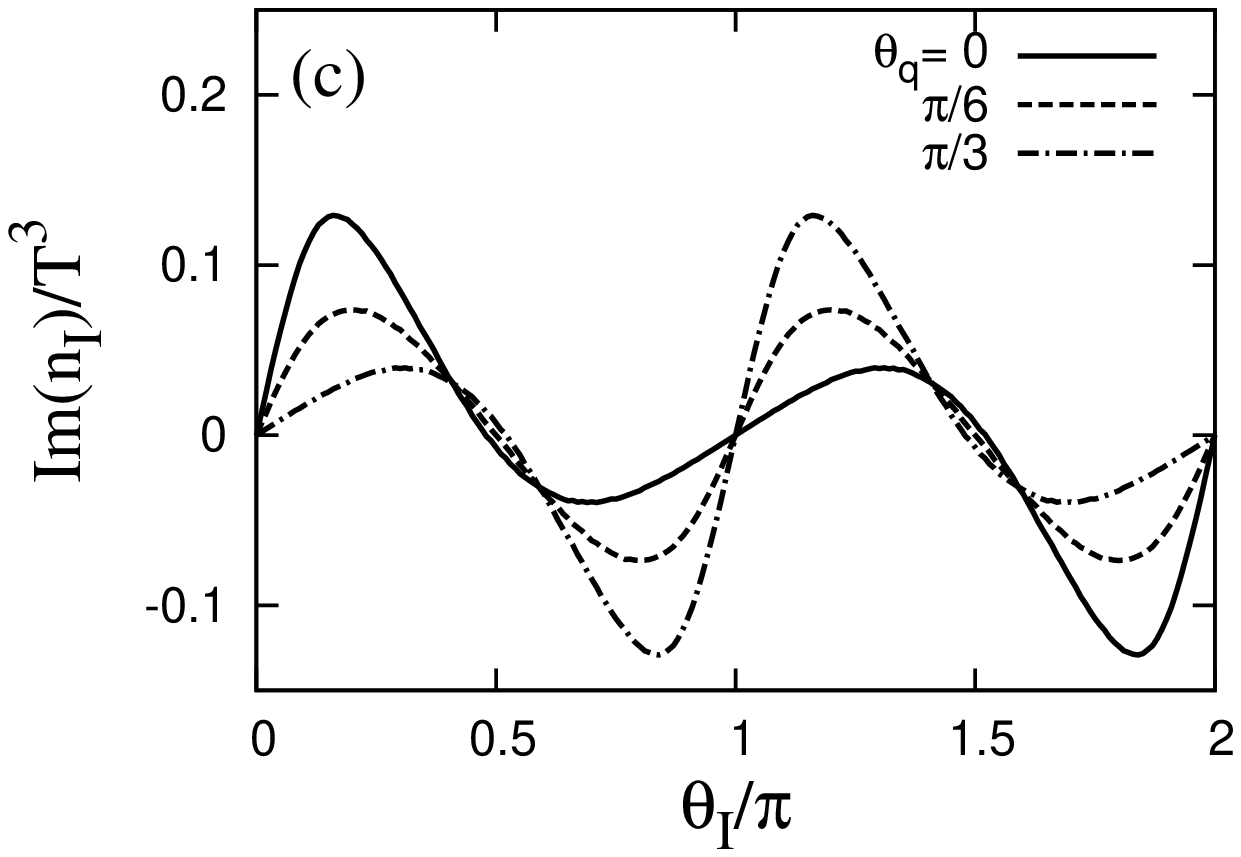}
\end{center}
\caption{$\theta_{\rm I}$ dependence of (a) $\Omega/T^4$, 
(b) ${\rm Im}[n_{\rm q}]/T^3$ and (c) ${\rm Im}[n_{\rm I}]/T^3$ at 
$T=175$~MeV. 
Three cases of $\theta_{\rm q}=0, \pi/6$ and $\pi/3$ are taken. 
In panel (b), the solid and dot-dashed lines agree with 
the $x$ axis. 
}
\label{175T-I}
\end{figure}

Figure~\ref{175T-I} presents $\theta_{\rm I}$ dependence of 
$\Omega/T^4$, 
${\rm Im}[n_{\rm q}]/T^3$ and ${\rm Im}[n_{\rm I}]/T^3$ 
at $\theta_{\rm q}=0, \pi/6$ and $\pi/3$ 
for the case of $T=175$MeV 
that is just above $T_{\rm c}=173$MeV and below $T_{\rm RW}$. 
The quantities $\Omega$ and ${\rm Im}[n_{\rm q}]$ are symmetric with respect 
to the axis $\theta_{\rm I}=\pi$, while ${\rm Im}[n_{\rm I}]$ is asymmetric 
with respect to the axis, as predicted 
by \eqref{pi-periodicity-2} - \eqref{pi-periodicity-4}. 
These are smooth everywhere in $\theta_{\rm I}$, 
and have a periodicity of $2\pi$ for all $\theta_{\rm q}$. 
For $\theta_{\rm q}=\pi/6$, $\Omega$, ${\rm Im}[n_{\rm I}]$ 
(${\rm Im}[n_{\rm q}]$) has a periodicity (anti-periodicity) 
of $\pi$ in $\theta_{\rm I}$, 
as expected from \eqref{pi/6-1} and \eqref{pi/6-2}. 
Below $T_{\rm RW}$, ${\rm Im}[n_{\rm q}]$ is smooth 
at any $\theta_{\rm q}$. Hence, $\theta_{\rm q}$-odd quantities like 
${\rm Im}[n_{\rm q}]$ are zero at $\theta_{\rm q}=0$ and $\pi/3$ mod $2\pi/3$. 
In panels (a) and (b), 
as a result of the property of \eqref{constant}, all curves almost meet 
at $\theta_{\rm I}=\pi/2$ and $3\pi /2$. 
As predicted by \eqref{shift}, in all panels,  
the dot-dashed curve for the case of $\theta_{\rm q}=\pi/3$ 
is obtained by shifting the solid one for 
the case of $\theta_{\rm q}=0$ by $\pi$ 
in the $\theta_{\rm I}$ direction.

\begin{figure}[htbp]
\begin{center}
 \includegraphics[width=0.30\textwidth,angle=-90]{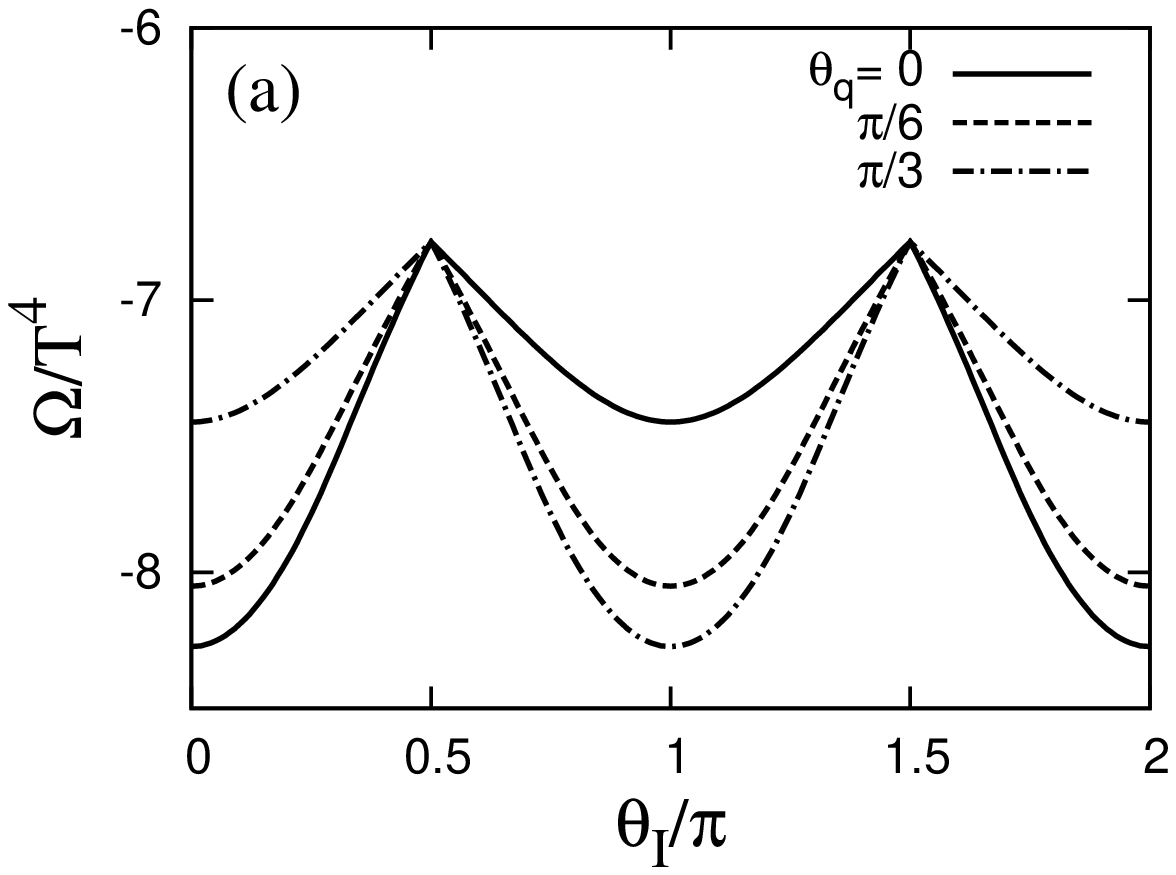}
 \includegraphics[width=0.30\textwidth,angle=-90]{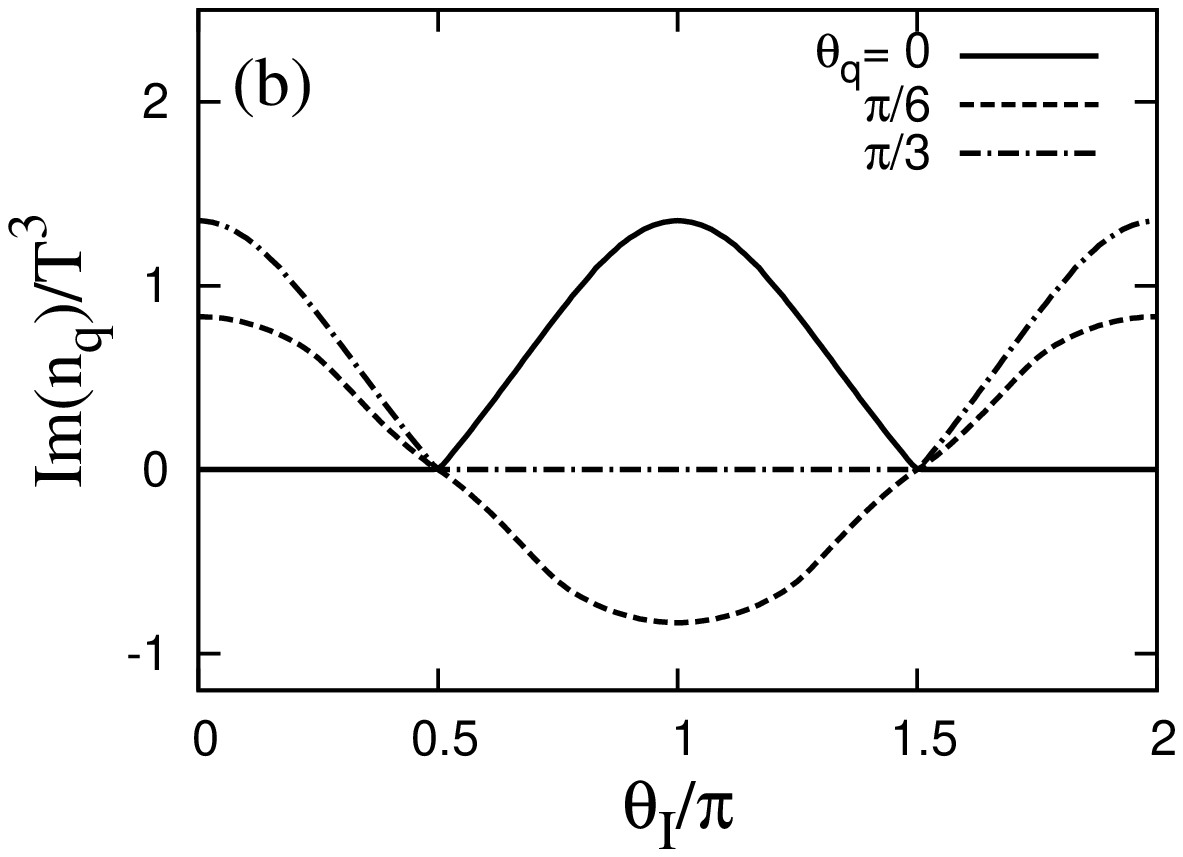}
 \includegraphics[width=0.30\textwidth,angle=-90]{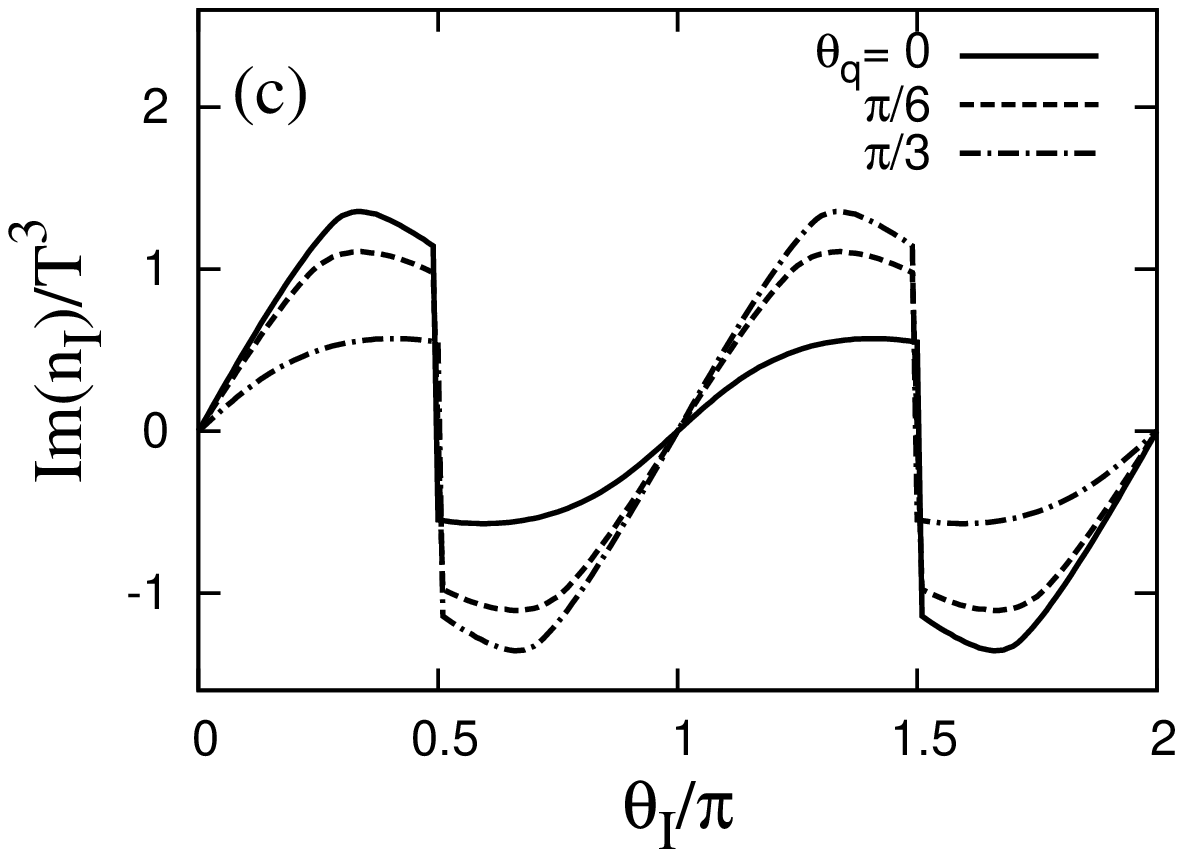}
\end{center}
\caption{$\theta_{\rm I}$ dependence of (a) $\Omega/T^4$, 
(b) ${\rm Im}[n_{\rm q}]/T^3$ and (c) ${\rm Im}[n_{\rm I}]/T^3$ at 
$T=250$~MeV. 
Three cases of $\theta_{\rm q}=-0, \pi/6$ and $\pi/3-0$ are taken. 
}
\label{250T-I}
\end{figure}

Figure \ref{250T-I} shows the same quantities as 
Fig.~\ref{175T-I}, but $T$ is taken to be 250~MeV as an example of 
temperature above $T_{\rm RW}$.
Again, $\Omega$ and ${\rm Im}[n_{\rm q}]$ are symmetric with respect 
to the axis $\theta_{\rm I}=\pi$, while ${\rm Im}[n_{\rm I}]$ is asymmetric 
with respect to the axis. All the quantities have a periodicity of $2\pi$ in 
$\theta_{\rm I}$ for all $\theta_{\rm q}$. 
For $\theta_{\rm q}=\pi/6$, $\Omega$ and ${\rm Im}[n_{\rm I}]$ 
have a periodicity of $\pi$ in $\theta_{\rm I}$, while 
${\rm Im}[n_{\rm q}]$ has an anti-periodicity of $\pi$ in $\theta_{\rm I}$. 
Hereafter, we consider 
the right-hand (left-hind) limit of $f(x)$ as $x$ approaches $a$ and denote 
it by $f(x)|_{x=a{\pm}0}$. 
As predicted by \eqref{shift}, 
the dot-dashed curve for the case of $\theta_{\rm q}=\pi/3-0$ 
is obtained by shifting the solid one for the 
case of $\theta_{\rm q}=-0$ by $\pi$ in the $\theta_{\rm I}$ direction. 
All curves almost meet at $\theta_{\rm I}=\pi/2$ and $3\pi/2$. 
$\theta_{\rm I}$-even quantities such as $\Omega$ and $n_{\rm q}$ have 
cusps at $\theta_{\rm q}=\pi/2$ mod $\pi$, so that $\theta_{\rm I}$-odd quantities such as $n_{\rm I}$ are discontinuous there. 
These singular behaviors represent the RW phase transition.

\subsection{Thermodynamics as a function of $\theta_{\rm q}$ and 
$\theta_{\rm I}$}

\begin{figure}[htbp]
\begin{center}
 \includegraphics[width=0.49\textwidth]{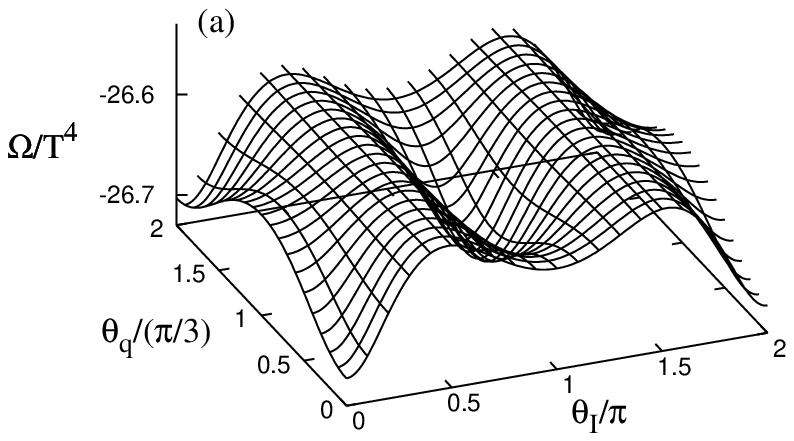}
 \includegraphics[width=0.49\textwidth]{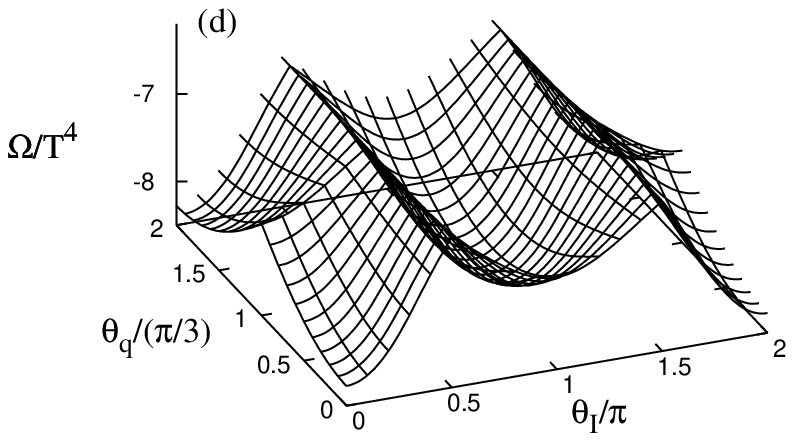}
 \includegraphics[width=0.49\textwidth]{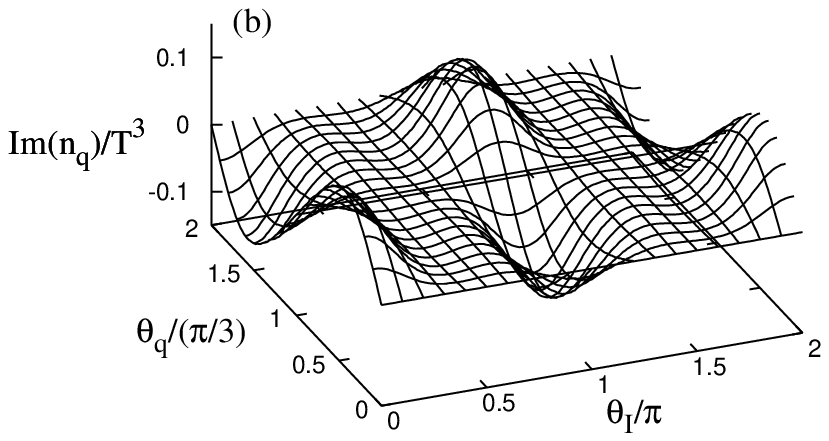}
 \includegraphics[width=0.49\textwidth]{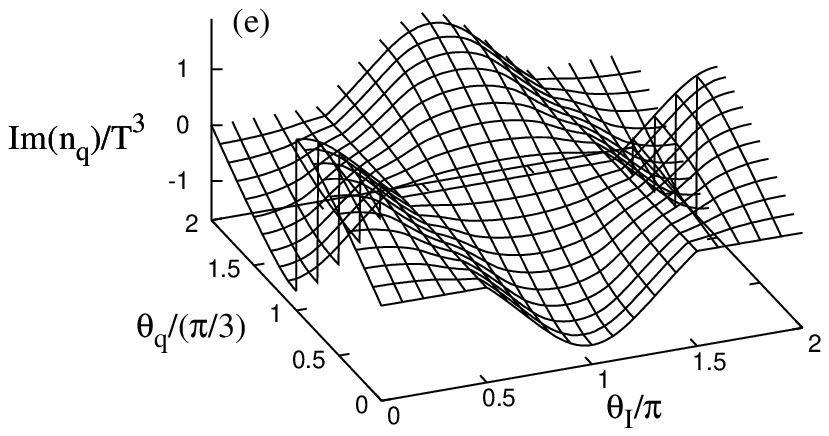}
 \includegraphics[width=0.49\textwidth]{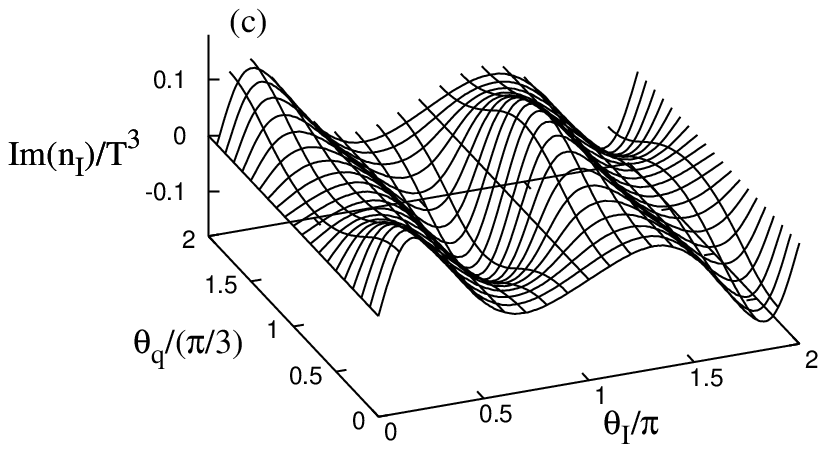}
 \includegraphics[width=0.49\textwidth]{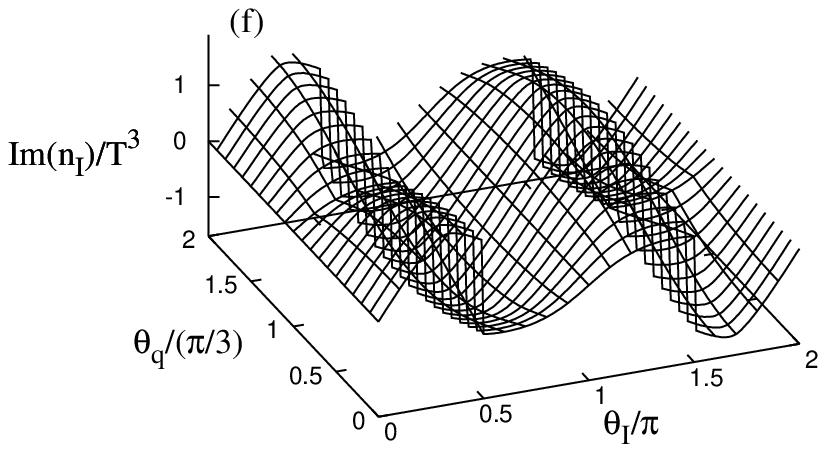}
\end{center}
\caption{$\Omega/T^4$, ${\rm Im}[n_{\rm q}]/T^3$ and 
${\rm Im}[n_{\rm I}]/T^3$ 
as a function of $\theta_{\rm q}$ and $\theta_{\rm I}$. 
Panels (a), (b) and (c) correspond to 175~MeV, while 
panels (d), (e) and (f) to 250~MeV.  
}
\label{Q-I}
\end{figure}

Figure \ref{Q-I} presents $\Omega/T^4$, ${\rm Im}[n_{\rm q}]/T^3$ and 
${\rm Im}[n_{\rm I}]/T^3$ as a function of $\theta_{\rm q}$ and 
$\theta_{\rm I}$ in the case of 
$T=175$ and 250~MeV. The symmetries \eqref{shift}-\eqref{pi/6-2} are 
seen as a bird's eye view. 
This result is consistent with LQCD ones~\cite{D'Elia-2}; 
in particular, discrete symmetry \eqref{pi/6-2} is clearly seen in 
the LQCD data. 
If the pion condensate is nonzero, the symmetries \eqref{shift}-\eqref{pi/6-2} 
break down, as shown in Sec.~\ref{PNJL}. Hence, the fact that LQCD has 
symmetry \eqref{shift} means 
that the pion condensation does not take place also 
in LQCD simulation.

\subsection{Comparison of PNJL results with LQCD results}

\begin{figure}[htbp]
\begin{center}
 \includegraphics[width=0.45\textwidth]{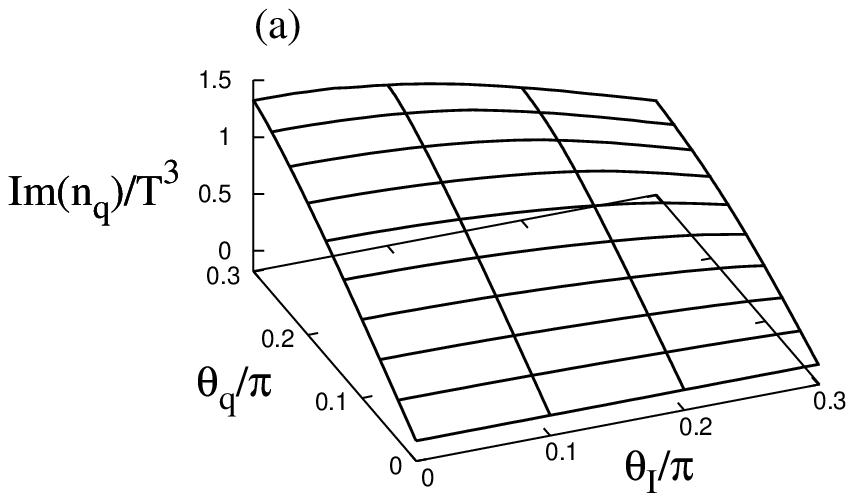}
 \includegraphics[width=0.45\textwidth]{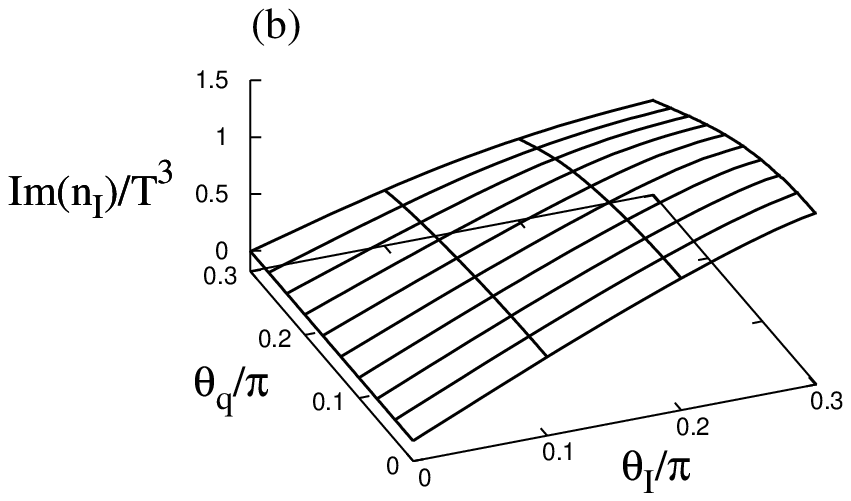}
\end{center}
\caption{(a) ${\rm Im}[n_{\rm q}]/T^3$ and (b) ${\rm Im}[n_{\rm I}]/T^3$
as a function fo 
$\theta_{\rm q}$ and $\theta_{\rm I}$ for the case of $T=250$MeV. }
\label{Q-I-2}
\end{figure}

LQCD data are available at temperatures below and above 
$T_{\rm RW}$ in Ref.~\cite{D'Elia-2}, where 
the lattice size is $16^3\times 4$ and the forth-rooted 
KS fermion is taken. 
The quark and isospin number densities, 
${\rm Im}[n_{\rm q}]/T^3$ and ${\rm Im}[n_{\rm I}]/T^3$, 
shown in Fig.~\ref{Q-I} 
are qualitatively consistent with the corresponding LQCD results presented 
in Figs. 2, 3, 9 and 10 of Ref.~\cite{D'Elia-2}.  
For the case of $T > T_{\rm RW}$, the consistency is more clearly seen 
in Fig.~\ref{Q-I-2} where 
${\rm Im}[n_{\rm q}]/T^3$ and ${\rm Im}[n_{\rm I}]/T^3$ 
are plotted in 
the same scale, $\theta_{\rm q}/\pi<0.3$ and $\theta_{\rm I}/\pi<0.3$, 
as Figs. 9 and 10 of Ref.~\cite{D'Elia-2}. 
In Ref.~\cite{D'Elia-2}, LQCD data on $n_{\rm q}$ and $n_{\rm I}$ 
are fitted by the hadron resonance gas (HRG) model~\cite{Karsch5} 
for the case of $T \le T_{\rm c}$, since the model is one of the 
most reliable models at $T < T_{\rm c}$ and also at $T=T_{\rm c}$
the model is successful in fitting 
the LQCD data by adding correction terms to it.
This makes more precise comparison possible for $T \le T_{\rm c}$.

In the HRG model, 
${\rm Im}[n_{\rm q}]$ and ${\rm Im}[n_{\rm I}]$ are obtained by 
sums of free-gas densities over kinds of particles~\cite{D'Elia-2}:
\begin{eqnarray}
{\rm Im}[n_{\rm q}]/T^3&=&\sum_{B, I\ge 0}3B~W_{B, I}(T)\bar{\delta}(I)
\sin(3B\theta_{\rm q})\cos(2I\theta_{\rm I}), 
\label{HRG-NQ}
\\
{\rm Im}[n_{\rm I}]/T^3&=&\sum_{B, I\ge 0}2I~W_{B, I}(T)\bar{\delta}(B)
\cos(3B\theta_{\rm q})\sin(2I\theta_{\rm I}), 
\label{HRG-NI}
\end{eqnarray}
where $\bar{\delta}(n)=1-\delta_{n,0}/2$ and $B$ ($I$) is 
the baryon (isospin) number of particle. 
The parameters are fitted to LQCD data in 
$\theta_{\rm q}-\theta_{\rm I}$ plane. The resultant values are summarized 
in Table II. 
\begin{table}[h]
\begin{center}
\begin{tabular}{ccccccccc}
\hline
$T$&$W_{0,1}$&$W_{0,2}$&$W_{1,1/2}$&$W_{1,3/2}$&$W_{1,5/2}$&$W_{1,7/2}$
&$W_{2,1}$&$W_{2,2}$
\\
\hline
$0.951T_{\rm c}$&~~~0.257~~~&~~~0.0106~~~&~~~0.0212~~~&~~~0.0265~~~&
~~~0.0009~~~&~~~0.0006~~~&~~~0.00090~~~&~~~...~~~\\
\hline
$T_{\rm c}$&~~~0.3214~~~&~~~0.0220~~~&~~~0.0344~~~&~~~0.0393~~~&~~~0.0042~~~&
~~~0.0015~~~&~~~0.0031~~~&~~~0.010~~~\\
\hline
\end{tabular}
\caption{ Summary of the parameter set of the HRG model in Table IV of
Ref.~\cite{D'Elia-2}. 
}
\end{center}
\end{table}

\begin{figure}[htbp]
\begin{center}
 \includegraphics[width=0.40\textwidth]{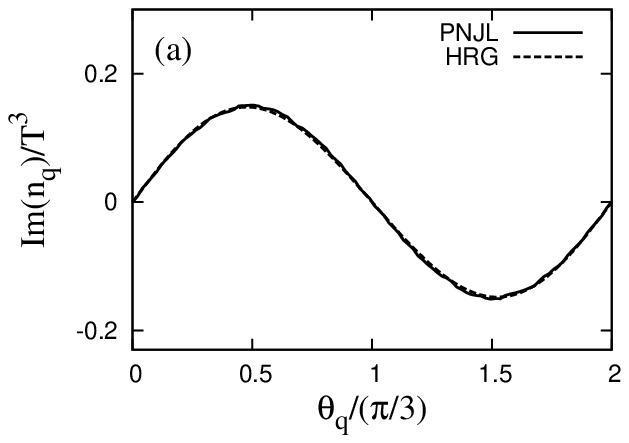}
 \includegraphics[width=0.40\textwidth]{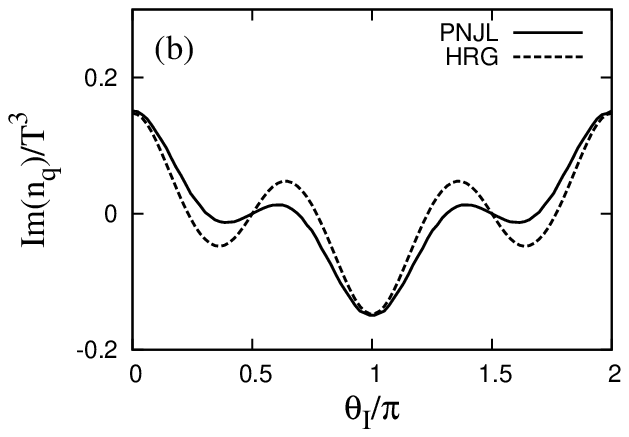}
 \includegraphics[width=0.40\textwidth]{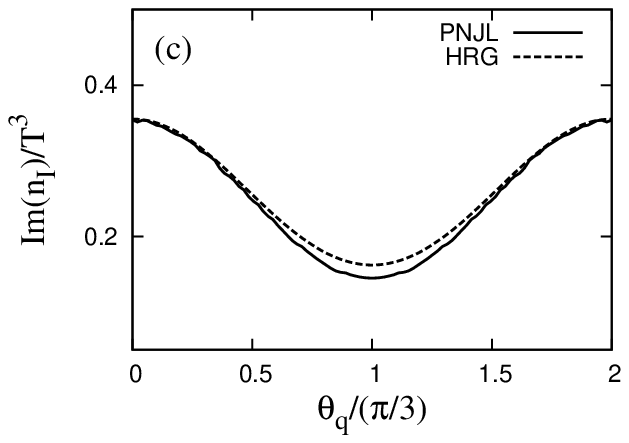}
 \includegraphics[width=0.40\textwidth]{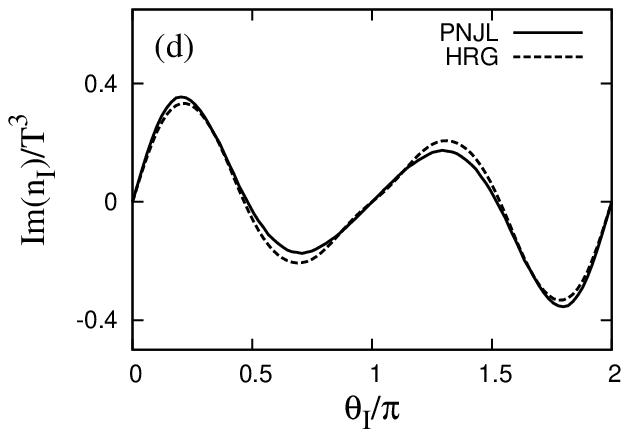}
\end{center}
\caption{Comparison of the PNJL model with 
the HRG model for 
${\rm Im}[n_{\rm q}]/T^3$ and ${\rm Im}[n_{\rm I}]/T^3$ 
at $T=0.951 T_{\rm c}=165$ MeV; 
(a) $\theta_{\rm q}$ dependence of 
${\rm Im}[n_{\rm q}]/T^3$ at $\theta_{\rm I}=0$, 
(b) $\theta_{\rm I}$ dependence of ${\rm Im}[n_{\rm q}]/T^3$ 
at $\theta_{\rm q}=\pi/6$, 
(c) $\theta_{\rm q}$ dependence of ${\rm Im}[n_{\rm I}]/T^3$ 
at $\theta_{\rm I}=\pi/5$ 
and 
(d) $\theta_{\rm I}$ dependence of ${\rm Im}[n_{\rm I}]/T^3$ 
at $\theta_{\rm q}=0$. 
The solid (dotted) lines denote the PNJL (HRG) results. 
The PNJL result is multiplied by 4 to fit the HRG result 
at $(\theta_{\rm q}, \theta_{\rm I})=(\pi/6, 0)$ in panels (a) 
and (b) and by 6.1 to fit the HRG result at 
$(\theta_{\rm q},\theta_{\rm I})=(0,\pi/5)$ in panels (c) and (d). 
}
\label{HRG}
\end{figure}

Figure~\ref{HRG} presents 
${\rm Im}[n_{\rm q}]/T^3$ and ${\rm Im}[n_{\rm I}]/T^3$ at $T=0.951 T_{\rm c}$. The solid (dotted) lines stand for the PNJL (HRG) results. 
In panels (a) and (b) where ${\rm Im}[n_{\rm q}]/T^3$ is plotted, 
the PNJL result is adjusted to the HRG result at 
$(\theta_{\rm q},\theta_{\rm I})=(\pi/6,0)$ by multiplying the PNJL result 
by 4. 
In panels (c) and (d) where ${\rm Im}[n_{\rm I}]/T^3$ is drawn, 
the PNJL result is fitted to the HRG result at 
$(\theta_{\rm q},\theta_{\rm I})=(0,\pi/5)$ 
by multiplying the PNJL result by 6.1. 
Oscillatory patterns of the HGM results are well reproduced by the 
PNJL model.

\begin{figure}[htbp]
\begin{center}
 \includegraphics[width=0.40\textwidth]{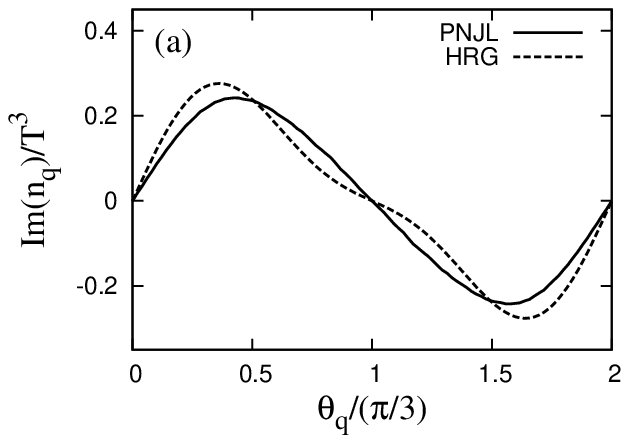}
 \includegraphics[width=0.40\textwidth]{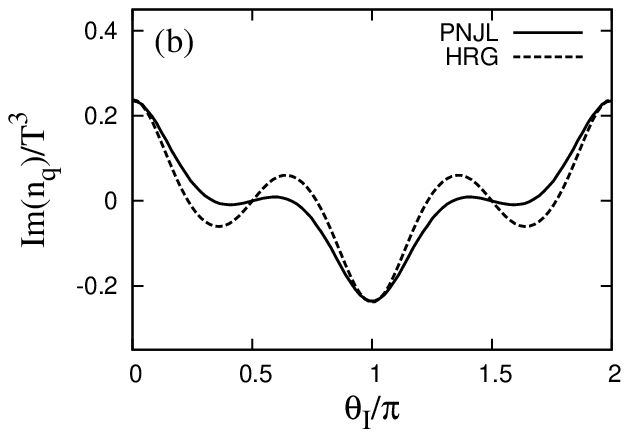}
 \includegraphics[width=0.40\textwidth]{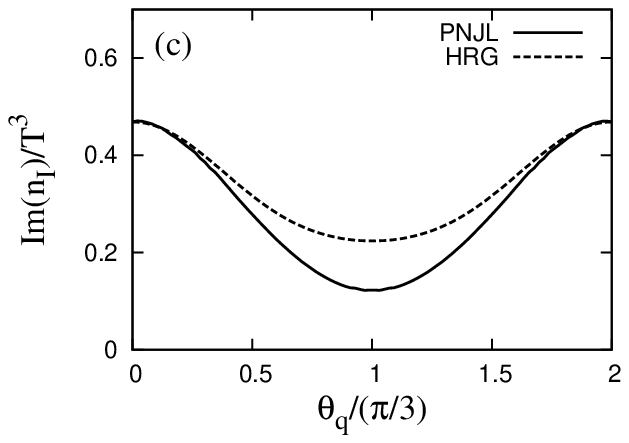}
 \includegraphics[width=0.40\textwidth]{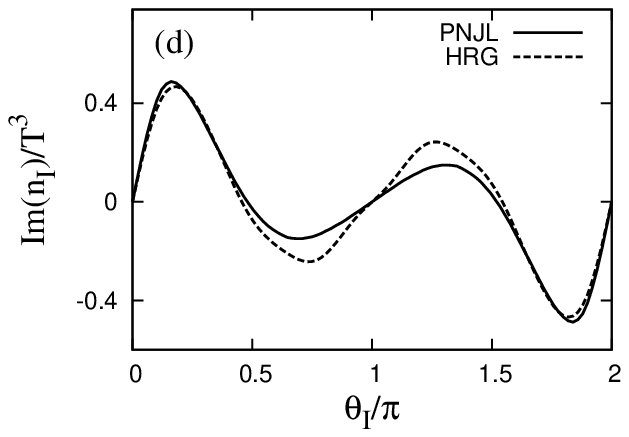}
\end{center}
\caption{Comparison of the PNJL model with 
the HRG model for 
${\rm Im}[n_{\rm q}]/T^3$ and ${\rm Im}[n_{\rm I}]/T^3$ 
at $T=175$ MeV ($\sim T_{\rm c}$); 
(a) $\theta_{\rm q}$ dependence of 
${\rm Im}[n_{\rm q}]/T^3$ at $\theta_{\rm I}=0$, 
(b) $\theta_{\rm I}$ dependence of ${\rm Im}[n_{\rm q}]/T^3$ 
at $\theta_{\rm q}=\pi/6$, 
(c) $\theta_{\rm q}$ dependence of ${\rm Im}[n_{\rm I}]/T^3$ 
at $\theta_{\rm I}=\pi/5$ 
and 
(d) $\theta_{\rm I}$ dependence of ${\rm Im}[n_{\rm I}]/T^3$ 
at $\theta_{\rm q}=0$. 
Definition of lines is the same as in Fig.~\ref{HRG}. 
The PNJL result is multiplied by 2.15 to fit the HRG result 
at $(\theta_{\rm q}, \theta_{\rm I})=(\pi/6, 0)$ in panels (a) 
and (b) and by 3.8 to fit the HRG result at 
$(\theta_{\rm q},\theta_{\rm I})=(0,\pi/5)$ in panels (c) and (d). 
}
\label{HRG-2}
\end{figure}

In Fig.~\ref{HRG-2}, the same analysis is made for $T=T_{\rm c}$. 
Again, 
the PNJL result is adjusted to the HRG result at 
$(\theta_{\rm q},\theta_{\rm I})=(\pi/6,0)$ by multiplying the PNJL result 
by 2.15 in panels (a) and (b) and at 
$(\theta_{\rm q},\theta_{\rm I})=(0,\pi/5)$ 
by multiplying the PNJL result by 3.8 in panels (c) and (d). 
Oscillatory patterns of the HRG results are reasonably reproduced by the 
PNJL model. 
Thus, the agreement between the two models becomes better in magnitude as 
$T$ increases. For the oscillatory pattern, the agreement is reasonably good 
at both $T=0.951 T_{\rm c}$ and $T_{\rm c}$.

\begin{figure}[htbp]
\begin{center}
 \includegraphics[width=0.40\textwidth]{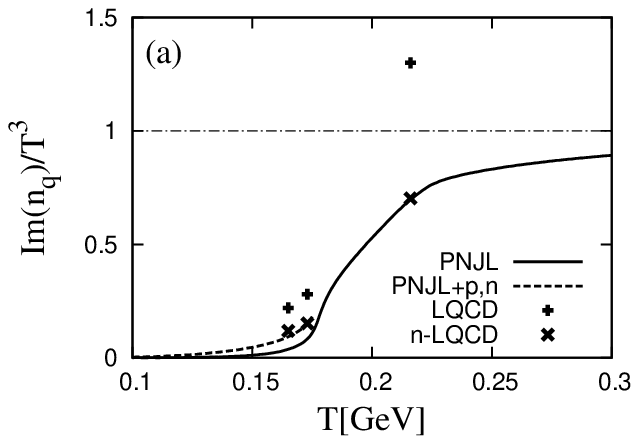}
 \includegraphics[width=0.40\textwidth]{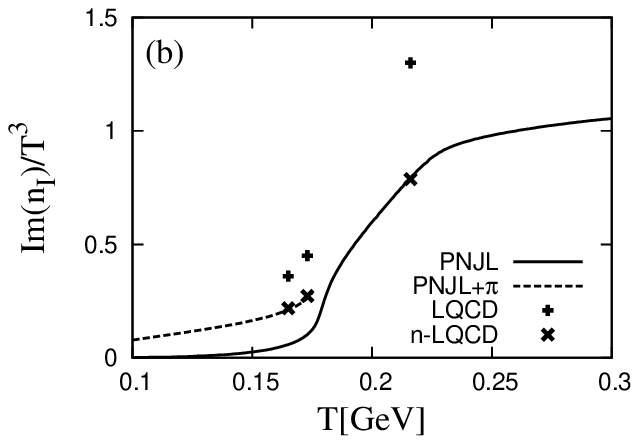}
\end{center}
\caption
{$T$ dependence of (a) ${\rm Im}(n_{\rm q})$ 
at $(\theta_{\rm q},\theta_{\rm I})=(\pi/6,0)$ 
and ${\rm Im}(n_{\rm I})$ at $(\theta_{\rm q},\theta_{\rm I})=(0,\pi/5)$. 
LQCD data are taken from \cite{D'Elia-2}.
The original values of LQCD data are plotted by plus symbols. 
The LQCD data are normalized so as to reproduce the PNJL result at 
$T=216$~MeV. The normalized LQCD (n-LQCD) data are shown by cross symbols. 
The dashed line is the result of the PNJL density plus 
the free-gas density; as a free particle we take nucleon for 
${\rm Im}(n_{\rm q})$ and pion for ${\rm Im}(n_{\rm I})$. 
The dot-dashed line represents ${\rm Im}(n_{\rm q})$ 
in the Stefan-Boltzmann limit. 
The LQCD result is multiplied by 0.53 and 0.60 to fit the PNJL result 
at $T=1.25T_{\rm c}$ in panels (a) and (b), respectively. 
}
\label{Nq-LQCD}
\end{figure}

The success of the PNJL model for the oscillatory pattern may indicate that 
the pattern is essentially controlled by discrete symmetries of 
\eqref{RW-periodicity}-\eqref{pi/3}. 
For magnitudes of 
${\rm Im}(n_{\rm q})$ and ${\rm Im}(n_{\rm I})$, meanwhile, 
the PNJL model underestimates LQCD results by a factor of $2 \sim 6$. 
Here we consider a possible origin of the discrepancy. 
In Fig. \ref{Nq-LQCD}(a),  
${\rm Im}(n_{\rm q})$ is plotted as a function of $T$ 
for the case of $(\theta_{\rm q},\theta_{\rm I})=(\pi/6,0)$. 
At $T=1.25T_{\rm c}=216$~MeV, LQCD data (plus symbol) is larger than 
the Stefan-Boltzmann high-$T$ limit (dot-dashed line).
Meanwhile,  
the PNJL result (solid curve) is smaller than the limit at $T=1.25T_{\rm c}$. 
The PNJL model is considered to be reliable 
above $T_{\rm c}$. Actually, for real quark chemical potential, 
the PNJL prediction on $n_{\rm q}$ is consistent with LQCD data~\cite{Ratti}. 
We then normalize the LQCD data so that the data at $T=1.25T_{\rm c}$ can 
agree with the PNJL result at $T=1.25T_{\rm c}$. The normalized data 
are shown by cross symbols. At $T=0.951T_{\rm c}=165$~MeV and 
$T_{\rm c}=173$~MeV, the PNJL result is smaller than 
the normalized data by a factor of about 2. 
This discrepancy is understandable as follows. 
Below $T_{\rm c}$, in general, hadronic excitations are important, 
but such an effect is not included in the mean-field approximation used 
in the present PNJL calculation. 
The ideal-gas model is considered to be good for $T<T_{\rm c}$ 
where hadrons have no decay modes. 
The ideal-gas model yields $W_{1,1/2} =0.0315$ for proton and neutron with 
physical masses. 
Substituting the value for $W_{1.1/2}$ in \eqref{HRG-NQ} 
and adding this correction to the original 
${\rm Im}(n_{\rm q})$ of the PNJL model, 
we have new ${\rm Im}(n_{\rm q})$. 
The new ${\rm Im}(n_{\rm q})$ is plotted by 
the dashed line up to $T_{\rm c}$. 
This line agrees with the normalized LQCD data 
at $T=165$ and 173~MeV.

The same analysis is possible for ${\rm Im}(n_{\rm I})$. 
Figure \ref{Nq-LQCD}(b) presents ${\rm Im}(n_{\rm I})$ as a function of $T$ 
for the case of $(\theta_{\rm q},\theta_{\rm I})=(0,\pi/5)$.
At $T=216$~MeV, LQCD data (plus symbol) is larger than the PNJL result by 
a factor of 1.5. Hence the data are normalized so that the data 
at $T=216$~MeV can reproduce the corresponding PNJL result. 
The data thus normalized are 
shown by cross symbols. At $T=T_{\rm c}=165$ and 173~MeV, 
the PNJL prediction underestimates the normalized LQCD data. 
Now the pion free-gas density is added to ${\rm Im}(n_{\rm I})$, 
where the pion mass is taken to be 280~MeV 
(the value of the LQCD calculation~\cite{D'Elia-2}). 
The new ${\rm Im}(n_{\rm I})$  is plotted by the dashed line up to 
$T_{\rm c}$. 
The new PNJL result 
agrees with LQCD data at $T=T_{\rm c}=165$ and 173~MeV.

As mentioned in Ref.~\cite{D'Elia-2}, 
the HRG model works well at $T<T_{\rm c}$, but not $T>T_{\rm c}$. 
At $T\sim T_{\rm c}$, corrections of a few percent to the model prediction 
are needed. This property is seen also in the PNJL result, as shown below. 

\begin{figure}[htbp]
\begin{center}
 \includegraphics[width=0.40\textwidth]{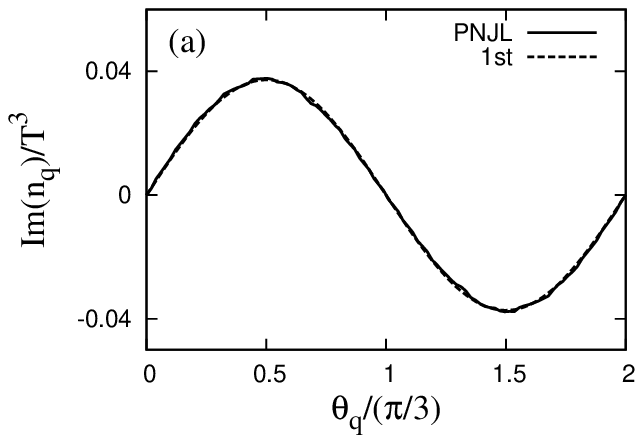}
 \includegraphics[width=0.40\textwidth]{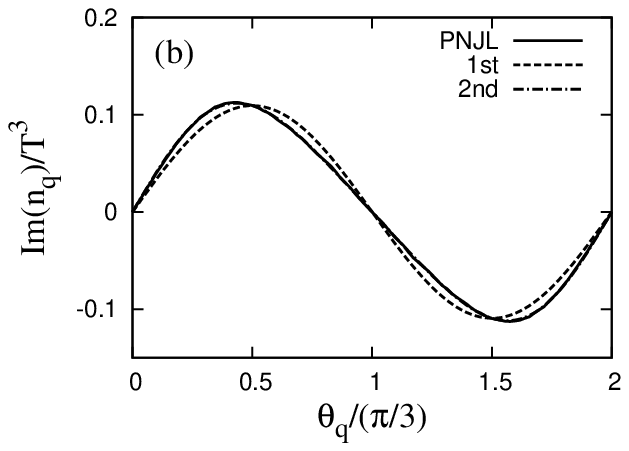}
 \includegraphics[width=0.40\textwidth]{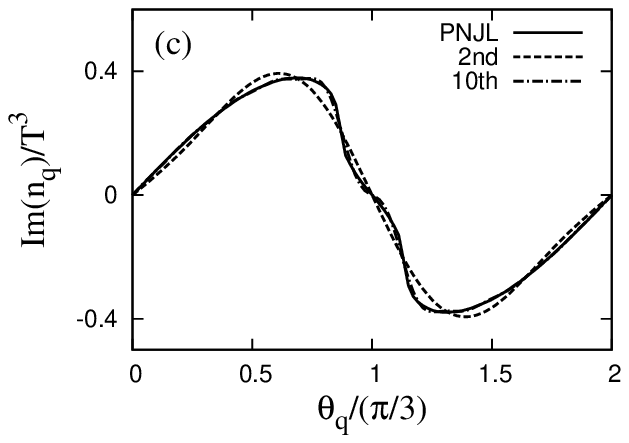}
\end{center}
\caption{
$\theta_{\rm q}$ dependence of $n_{\rm q}$ and 
$n_{\rm q}(k_{\rm max})$ for (a) $T=165$~MeV$<T_{\rm c}$, 
(b) $T=175$~MeV$\sim T_{\rm c}$ and (c) $T=185$~MeV$>T_{\rm c}$.
The case of $\theta_{\rm I}=0$ is taken. 
}
\label{sin}
\end{figure}

Noting that $n_{\rm q}$ is $\theta_{\rm q}$-odd, we can find from 
\eqref{k-expansion} that 
\begin{eqnarray}
n_{\rm q}=\sum_{k=0}^{\infty} a_k\sin(3k\theta_{\rm q}). 
\label{n-sin}
\end{eqnarray}
The $a_{k}$ terms with $k > 2$ correspond to corrections 
to the HRG model. 
Now, we introduce a partial sum 
\begin{eqnarray}
n_{\rm q}(k_{\rm max})=\sum_k^{k_{\rm max}} a_k\sin(3k\theta_{\rm q}), 
\label{n-sin-2}
\end{eqnarray}
where the $a_k$ are evaluated from $n_{\rm q}$ calculated with 
the PNJL model. 
Figure~\ref{sin} shows $\theta_{\rm q}$ dependence of $n_{\rm q}$ and 
$n_{\rm q}(k_{\rm max})$ with some values of $k_{\rm max}$, where 
the case of $\theta_{\rm I}=0$ is taken. 
Panels (a), (b) and (c) correspond to the cases 
of $T=165$~MeV ($< T_{\rm c}$), 175~MeV 
($\sim T_{\rm c}$) and 185~MeV ($>T_{\rm c}$), respectively. 
The PNJL result, $n_{\rm q}=n_{\rm q}(k_{\rm max}=\infty)$, 
is well approximated by 
$n_{\rm q}(k_{\rm max}=1)$ for $T<T_{\rm c}$, 
$n_{\rm q}(k_{\rm max}=2)$ for $T \sim T_{\rm c}$
and $n_{\rm q}(k_{\rm max}=10)$ for $T>T_{\rm c}$, as expected. 
The relative value $a_2/a_1$ at $T\sim T_{\rm c}$ 
is 0.11 for the PNJL result, while 0.12 for the LQCD result. 
Thus, the PNJL result is consistent with the LQCD result.

\subsection{Phase diagram in $\mu_{\rm I}$-$T$ plane}
\label{I-T-plane}

In this subsection, the phase diagram is explored mainly 
in $\mu_{\rm I}$-$T$ plane, since 
the phase diagram in $\mu_{\rm q}$-$T$ plane has 
already been analyzed for the case of $\mu_{\rm I}=0$ 
in Refs.~\cite{Sakai1,Sakai2,Kouno}.

\begin{figure}[htbp]
\begin{center}
 \includegraphics[width=0.50\textwidth]{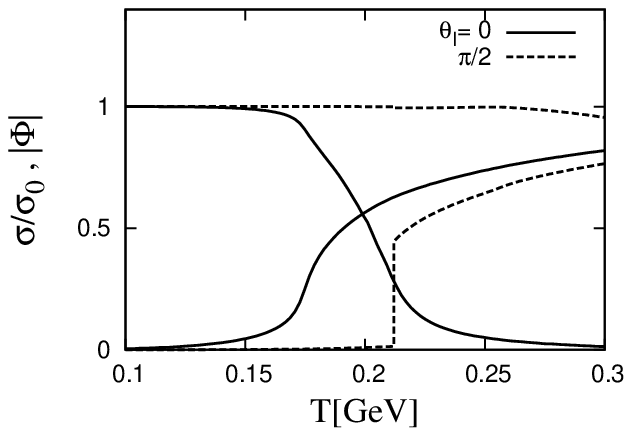}
\end{center}
\caption{$T$ dependence of $|\Phi|$ and $\sigma$ 
normalized by the value $\sigma_0$ at $T=\mu_{\rm q}=\mu_{\rm I}=0$; 
as $T$ increases, $|\Phi|$ increases but $\sigma$ decreases. 
The solid (dashed) curves correspond to the case of $\theta_{\rm I}=0$ 
($\pi/2$).  }
\label{Q00-T}
\end{figure}

Figure~\ref{Q00-T} presents $T$ dependence of 
the absolute value $|\Phi|$ and the chiral condensate $\sigma$ 
for two cases of $\mu_{\rm I}=0$ and $\pi/2$, 
where $\sigma$ is normalized by the value $\sigma_0$ 
at $T=\mu_{\rm q}=\mu_{\rm I}=0$. 
Note that $|\Phi|$ is an increasing function of $T$, while 
the normalized $\sigma$ is a decreasing function of $T$. 
When $\theta_{\rm I}=0$, 
both the chiral and the deconfinement transition are 
crossover, as represented by the solid curves. 
Their pseudo-critical temperatures are 
$T_{\rm c}^{\chi}=216$~MeV for the chiral transition and 
$T_{\rm c}^{\rm conf}=173$~MeV 
for the deconfinement transition in the present PNJL calculation, 
while $T_{\rm c}^{\chi} \approx T_{\rm c}^{\rm conf} =173 \pm 8$~MeV 
in LQCD calculation~\cite{Karsch}. 
Thus, the correlation between the two transitions is weaker 
in the present PNJL calculation than in the LQCD simulation. 
In the previous paper~\cite{Sakai2}, therefore, we 
introduced the scalar-type eight-quark interaction 
in the PNJL calculation 
in order to solve this problem; actually, 
$T_{\rm c}^{\chi} \approx T_{\rm c}^{\rm conf}=173 \pm 8$~MeV 
in the PNJL calculation with the scalar-type eight-quark interaction.

For $\theta_{\rm I}=\pi/2$, as denoted by the dashed curves 
in Fig.~\ref{Q00-T},    
the deconfinement phase transition becomes 
first order, while the chiral condensate hardly depends on $T$. 
As shown in \eqref{constant},  
the u-quark loop contribution to $\Omega$ is nearly 
canceled out by the d-quark one, when $\theta_{\rm I}=\pi/2$. 
As a consequence of 
this cancellation in $\Omega$, 
$\sigma$ has a weak $T$ dependence, while 
$T$ dependence of $\Phi$ is controlled 
by the pure gauge part ${\cal U}$. 
The potential ${\cal U}$ breaks the center symmetry {\it spontaneously}, 
when $T \ge T_0 =212$~MeV; 
as shown in Ref.~\cite{Kouno}, ${\cal U}$ has two local minima at $|\Phi|=0$ 
and $\sim 0.45$ for $T$ near $T_0$, 
and the local minimum at $|\Phi|=0$ is deeper than 
the other only when $T < T_0$. 
Eventually, $T_{\rm c}^{\rm conf}$ nearly agrees with $T_0$ and hence 
becomes much smaller than $T_{\rm c}^{\chi}$ 
in the present PNJL calculation with no the eight-quark interaction; 
i.e., $T_{\rm c}^{\rm conf}=212$~MeV and $T_{\rm c}^{\chi}=455$~MeV.
The difference between $T_{\rm c}^{\chi}$ and $T_{\rm c}^{\rm conf}$ 
is still large, even if the eight-quark interaction is taken into account; 
i.e., $T_{\rm c}^{\rm conf}=212$~MeV 
and $T_{\rm c}^{\chi}=405$~MeV.

Since two-flavor LQCD data are not available at $\theta_{\rm I}=\pi/2$, 
it is not clear whether the large difference is realistic. 
However, it should be noted that 
LQCD data at $\theta_{\rm I}=\pi/2$ are available 
in the 8-flavor case~\cite{Cea}. The data show that 
the chiral and deconfinement transitions are first order and 
$T_{\rm c}^{\chi} \approx T_{\rm c}^{\rm conf}$. 
Unfortunately, it is not straightforward 
to apply the PNJL model to the 8-flavor system, 
since the LQCD data do not present the pion mass and the pion decay constant 
and hence we cannot determine the parameters of the PNJL model. 
If the PNJL calculation is done without changing the parameters 
from the 2-flavor case, the calculation shows 
$T_{\rm c}^{\chi} \gg T_{\rm c}^{\rm conf}$ and therefore cannot 
reproduce the LQCD data. 
This disagreement of the PNJL result with the LQCD data 
is originated in the fact that 
the correlation between $\sigma$ and $\Phi$ is weak in the PNJL model. 
This suggests that $\Omega$ of the PNJL model 
should have a direct coupling term such as $\sigma \Phi \Phi^*$. 
Thus, the nature of the coincidence between the chiral and deconfinement 
transitions in LQCD, 
or the origin of the direct-coupling term between $\sigma$ and $\Phi$ 
in the PNJL model, is an interesting subject as a future work.

\begin{figure}[htbp]
\begin{center}
 \includegraphics[width=0.30\textwidth,angle=-90]{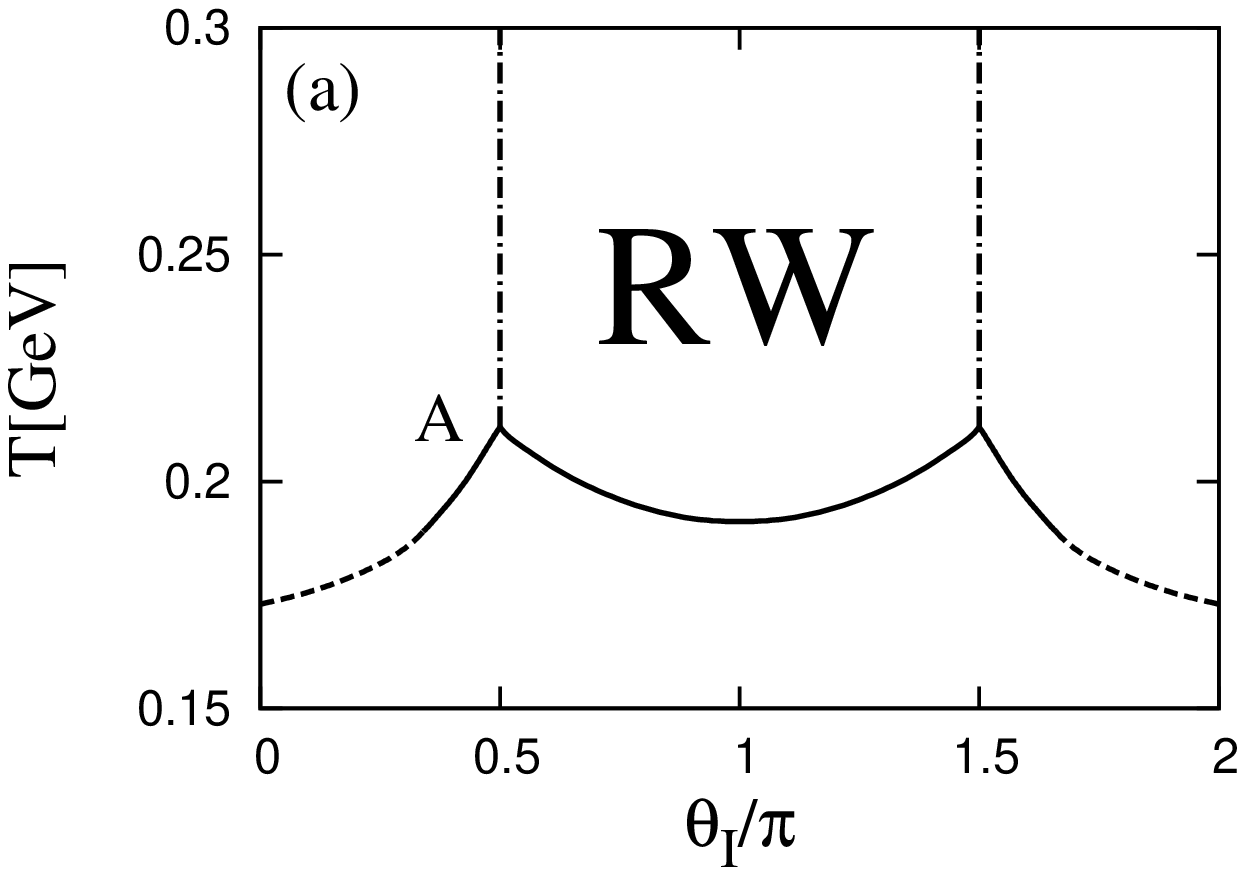}
 \includegraphics[width=0.30\textwidth,angle=-90]{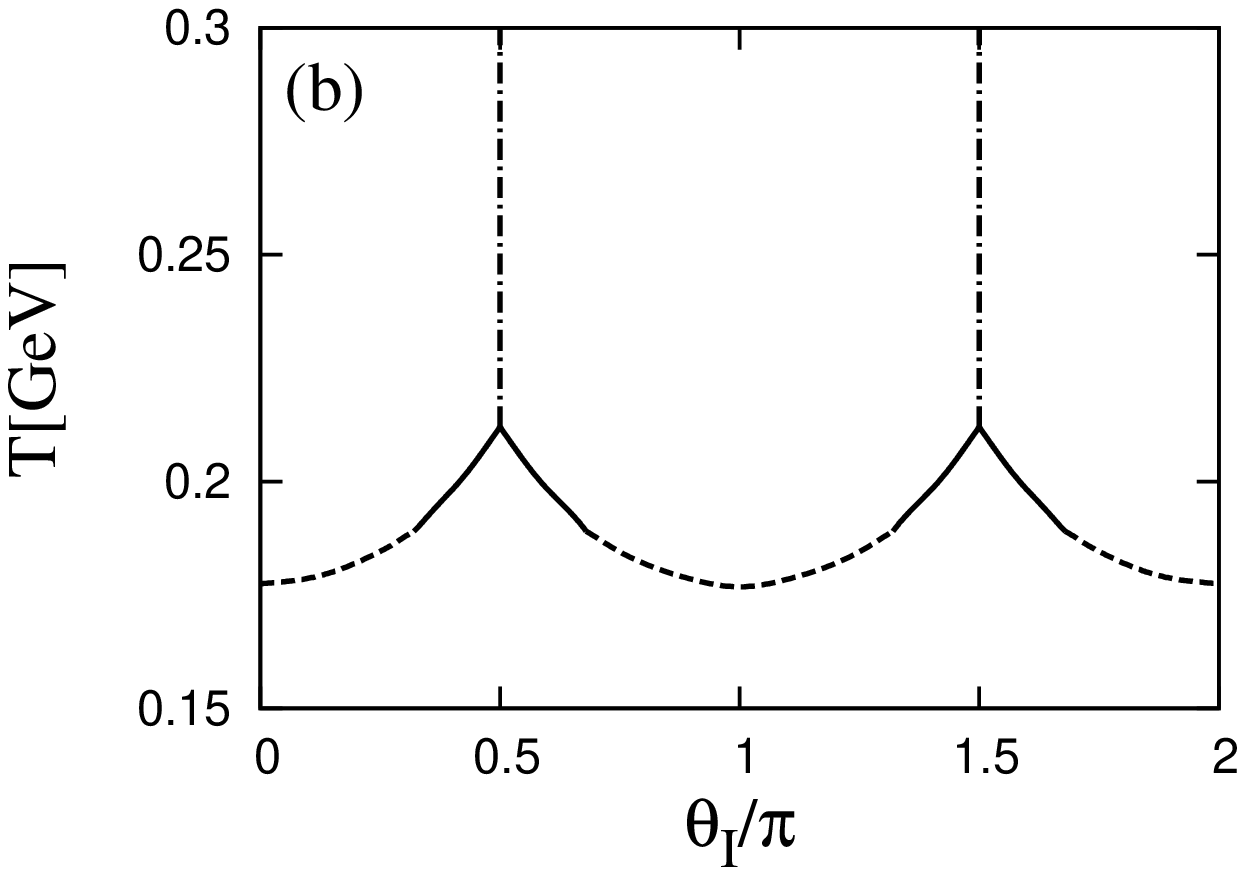}
 \includegraphics[width=0.30\textwidth,angle=-90]{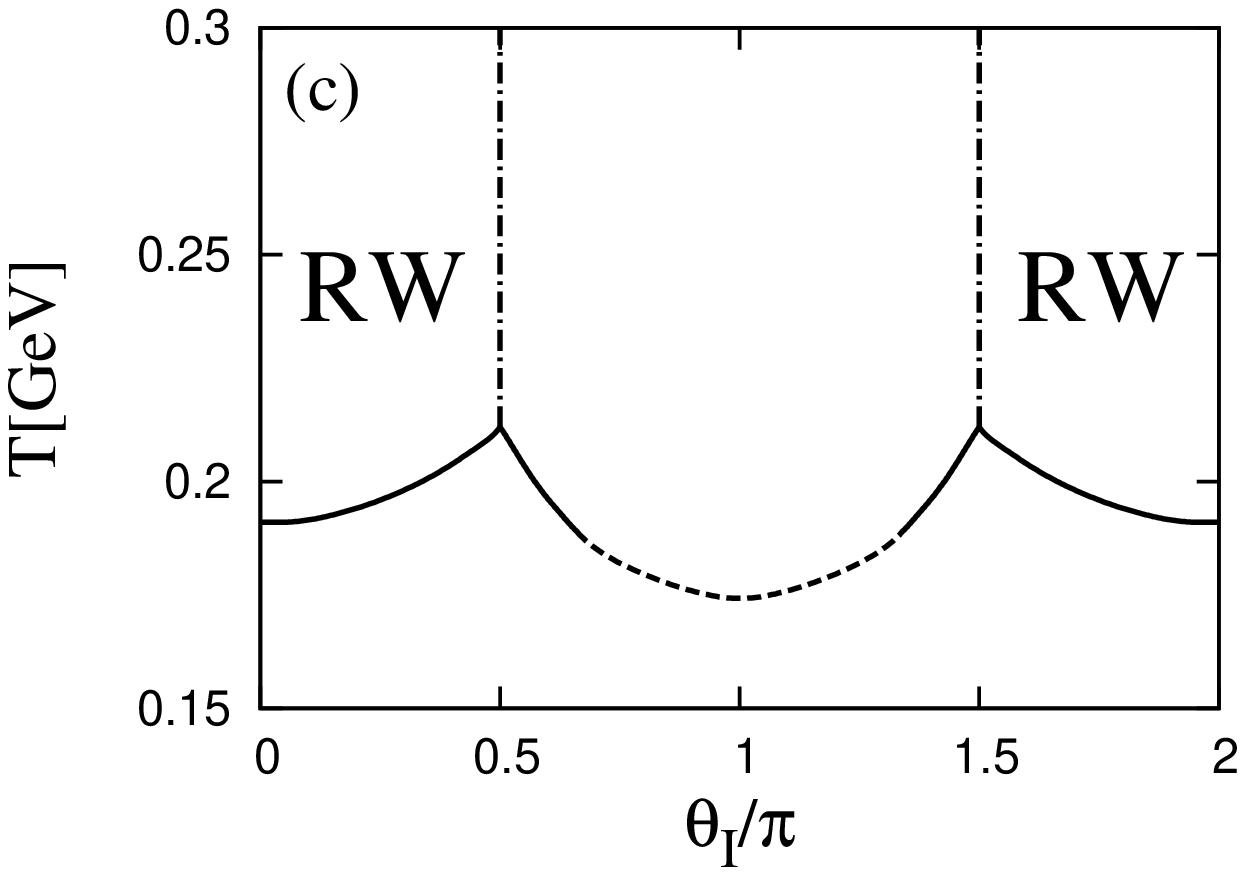}
\end{center}
\caption{Phase diagram of the deconfinement phase transition 
in $\theta_{\rm I}$-$T$ plane. Panels (a), (b) and (c) 
are the cases of 
$\theta_{\rm q}=0, \pi/6$ and $\pi/3$ respectively.
The first-order (crossover) transition is denoted by the solid (dashed) 
curves. The area labeled by ``RW" between the two dot-dashed lines represents 
the region in which the RW phase transition takes place. 
Point A is located at 
($T_{\rm A}, \theta_{\rm A})=(212~{\rm MeV}, 0.494\pi)$. 
}
\label{Q00-PD}
\end{figure}

Figure~\ref{Q00-PD} shows the phase diagram of the deconfinement phase 
transition in $\theta_{\rm I}$-$T$ plane, where panels 
(a), (b) and (c) correspond to three cases 
of $\theta_{\rm q}=0, \pi/6$ and $\pi/3$, respectively. 
The solid curves denote the first-order phase transition, while the dashed 
lines stand for the crossover transition. 
Near $\theta_{\rm I}=\pi/2$ mod $\pi$, 
the deconfinement phase transition are first order in all the cases. 
Near $\theta_{\rm I}=\pi$ mod $\pi$, 
the deconfinement phase transition is first order 
when $\theta_{\rm q}=0$, but crossover 
when $\theta_{\rm q}=\pi/6$ and $\pi/3$. 
The RW phase transition occurs in the area labeled by ``RW" 
between the two dot-dashed lines. 
The dot-dashed line is a boundary of the area and is 
called ``the RW-like transition line " 
in Ref.~\cite{Cea}. It is a nearly-vertical line 
starting from point A and is expressed 
as $\theta_{\rm I}=\pi /2-\delta (T)$ where $\delta(T)$ is defined in 
\eqref{delta}. 
Point A is located at ($T_{\rm A}, \theta_{\rm A})=(1.23T_{\rm c}, 0.494\pi)$ 
in the present 2-flavor analysis, while 8-flavor LQCD data~\cite{Cea}  
show ($T_{\rm A}, \theta_{\rm A})=(1.2T_{\rm c}, 0.48\pi)$. 
Thus, the present result seems to be consistent with the LQCD data.

\begin{figure}[htbp]
\begin{center}
 \includegraphics[width=0.40\textwidth]{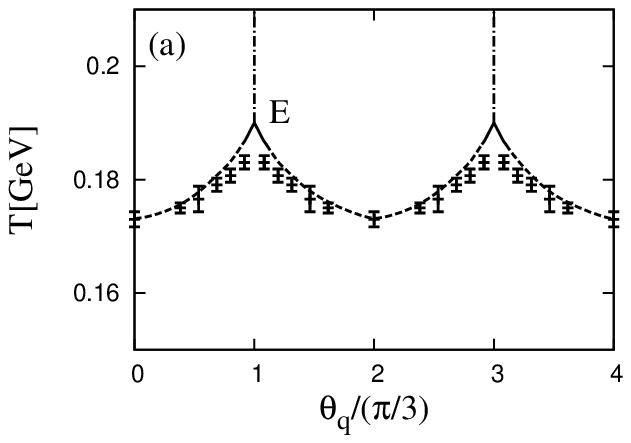}
 \includegraphics[width=0.40\textwidth]{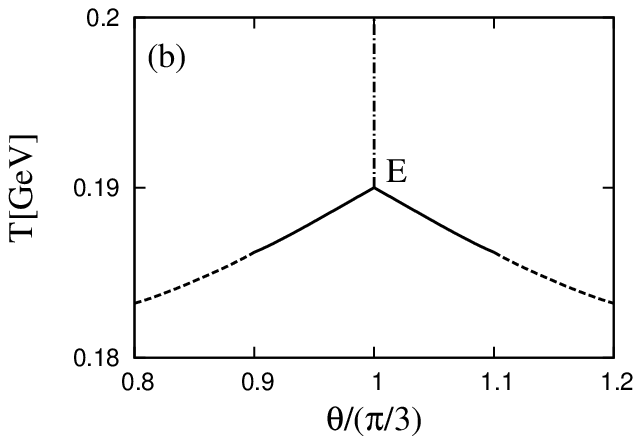}
\end{center}
\caption{Phase diagram of the deconfinement and the RW phase transition 
in $\theta_{\rm q}$-$T$ plane at $\theta_{\rm I}=0$. 
The solid lines stand for the first-order deconfinement transition,
while the dashed lines denote the crossover deconfinement transition. 
The RW transition is denoted by the dot-dashed curve. Point E is an 
endpoint of the RW transition. 
In panel (b), the phase structure near point E is magnified. 
Lattice data are taken from Ref.~\cite{Chen}; the pseudocritical 
temperature at $\theta_{\rm q}=0$ is assumed to be 173~MeV determined from 
LQCD calculation of Ref.~\cite{Karsch}. 
}
\label{PD-Q}
\end{figure}

\begin{figure}[htbp]
\begin{center}
 \includegraphics[width=0.40\textwidth]{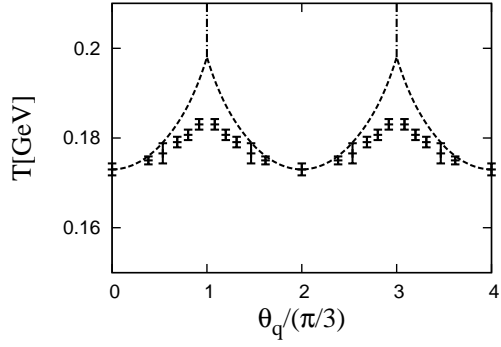}
\end{center}
\caption{
Phase diagram of the deconfinement and the RW phase transition 
in $\theta_{\rm q}$-$T$ plane at $\theta_{\rm I}=0$. 
Here, the Polyakov potential of F-type~\cite{Fukushima2} 
is taken in the PNJL calculation. See the figure caption of Fig.~\ref{PD-Q}
for definition of lines and lattice data. 
}
\label{PD-Q-2}
\end{figure}

\begin{figure}[htbp]
\begin{center}
 \includegraphics[width=0.40\textwidth]{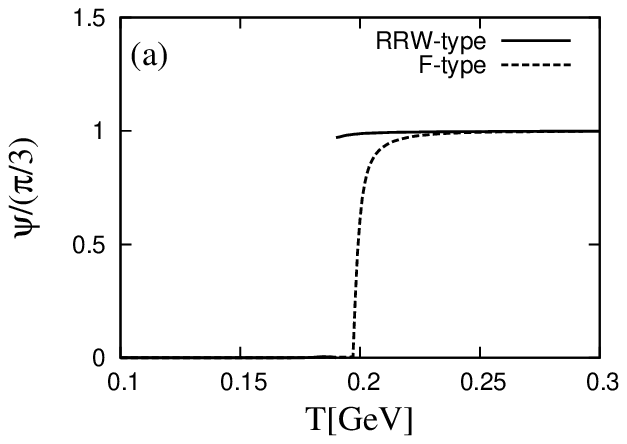}
 \includegraphics[width=0.40\textwidth]{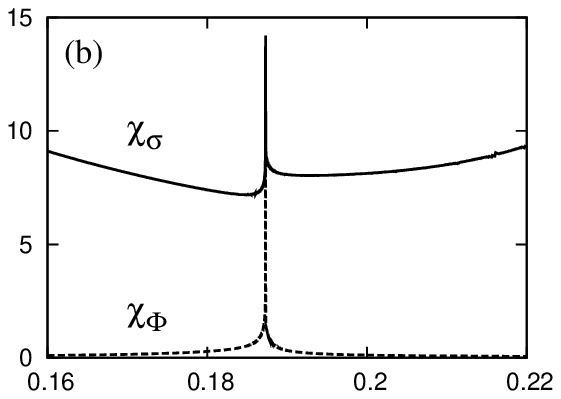}
\end{center}
\caption{
(a) $T$ dependence of phase $\psi$ 
of the modified Polyakov-loop $\Psi_f$ 
at $\theta_{\rm q}=\pi/3$ and $\theta_{\rm I}=0$. 
The solid line shows the result of RRW-type $\cal U$~\cite{Rossner}, 
while the dashed line corresponds to the result of 
F-type $\cal U$~\cite{Fukushima2}. 
(b) $T$ dependence of 
the chiral and Polyakov-loop susceptibilities, $\chi_{\sigma}$ and 
$\chi_{\Phi}$, at 
$\theta_{\rm q}=0.93 \times \pi/3$ and $\theta_{\rm I}=0$. 
}
\label{T-Poly-3}
\end{figure}

Figure~\ref{PD-Q} shows the phase diagram of the deconfinement and the 
RW phase transition in $\theta_{\rm q}$-$T$ plane 
at $\theta_{\rm I}=0$. 
The solid lines represent the first-order deconfinement transition, 
while the dashed lines do the crossover deconfinement transition. 
The dot-dashed lines stand for the RW transition line, while 
point E denotes an endpoint of the RW transition. 
In panel (a), the present PNJL model reproduces 
LQCD data~\cite{Chen} at finite $\theta_{\rm q}$. 
The phase diagram near the RW endpoint (point E) is magnified in panel (b). 
Thus, the RW endpoint is first order in the present 
PNJL calculation with RRW-type ${\cal U}$~\cite{Rossner}; 
detailed analyses will be made later in Fig.~\ref{T-Poly-3}. 
However, it was second order in the previous PNJL calculation~\cite{Kouno} 
with F-type ${\cal U}$~\cite{Fukushima2} in which 
a form inspired by a strong coupling QCD was taken for ${\cal U}$. 
Thus, the order of the deconfinement phase transition near 
the RW endpoint strongly depends on ${\cal U}$ taken. 
For comparison, the previous PNJL result is plotted together with 
LQCD data in Fig.~\ref{PD-Q-2}. 
Thus, the present calculation gives better agreement with 
LQCD data than the previous one. In this sense, 
the present PNJL calculation is more reliable. 
The result of the present PNJL calculation is consistent 
with a latest LQCD result~\cite{D'Elia-3} 
in which the order of the RW phase transition at point E 
is first order for small quark mass, although it is second order 
for heavy quark mass. 

Finally, the behavior of the RW transition near endpoint E is 
analyzed more explicitly. 
Figure~\ref{T-Poly-3}(a) shows $T$ dependence of phase $\psi$ of the 
modified Polyakov-loop $\Psi_f$ at $\theta_{\rm q}=\pi/3$ and $\theta_{\rm I}=0$. 
The solid line shows the PNJL prediction with 
RRW-type ${\cal U}$, while the dashed line corresponds to the result of 
F-type ${\cal U}$. 
The phase $\psi$ is an order parameter of the RW phase transition~\cite{Kouno}. 
Obviously, the RW phase transition at endpoint E is first order for RRW-type ${\cal U}$, but second order for F-type ${\cal U}$. 
As shown in Fig.~\ref{PD-Q} (b), there is 
a meeting point of the solid and dashed lines 
at $T=0.187$~MeV, $\theta_{\rm q}=0.93 \times \pi/3$ 
and $\theta_{\rm I}=0$. 
This is a critical endpoint of the 
deconfinement phase transition by definition. 
Figure~\ref{T-Poly-3}(b) presents 
the chiral and Polyakov-loop susceptibilities, $\chi_{\sigma}$ and 
$\chi_{\Phi}$, as a function of $T$ 
at $\theta_{\rm q}=0.93 \times \pi/3$ and $\theta_{\rm I}=0$, 
where RRW-type ${\cal U}$ is taken. 
The susceptibilities are divergent at the critical endpoint. Hence, 
the chiral and deconfinement transitions are second order at 
the critical endpoint.

\section{Summary}
\label{Summary}

We have explored the phase diagram of two-flavor QCD 
at imaginary quark-number and isospin chemical potentials, 
$\mu_{\rm q}=iT\theta_{\rm q}$ and $\mu_{\rm iso}=iT\theta_{\rm iso}$. 
At imaginary $\mu_{\rm iso}$, the pion condensation does not 
take place. The QCD vacuum is then $I_3$ symmetric. 
As a consequence, at imaginary $\mu_{\rm iso}$ and $\mu_{\rm q}$, 
the partition function (the thermodynamic potential) 
has discrete symmetries \eqref{RW-periodicity}-\eqref{pi-periodicity}
that are not present at 
real $\mu_{\rm iso}$ and $\mu_{\rm q}$. 
The PNJL model possesses all the discrete symmetries, and hence 
the PNJL results are qualitatively consistent with 
LQCD data presented very lately~\cite{Cea,D'Elia-2}. 
In particular, LQCD data~\cite{D'Elia-2} 
have symmetry \eqref{pi/6-2} derived from 
\eqref{RW-periodicity}-\eqref{pi/3}. 
This indicates that the pion condensation does not occur in the 
LQCD calculation.

A quantitative comparison of the PNJL model with 
LQCD data~\cite{Cea,D'Elia-2} is made at $T \le T_{\rm c} $ 
by using the hadron resonance gas (HRG) model that 
can reproduce the LQCD data there. 
As for ${\rm Im}[n_{\rm q}]$ and ${\rm Im}[n_{\rm I}]$, 
the PNJL result underestimates the HRG result in magnitude, but for 
$\theta_{\rm q}$ and $\theta_{\rm I}$ dependences 
the agreement between the two is reasonably good. 
Thus, the PNJL model is useful 
at imaginary $\mu_{\rm iso}$ and $\mu_{\rm q}$.

The PNJL model predicts that 
the RW phase transition occurs at $\theta_{\rm q}=\pi/3$ mod 
$2\pi/3$ when $-\pi/2-\delta(T)<\theta_{\rm I}
=\theta_{\rm iso}/2<\pi/2+\delta(T)$, 
while at $\theta_{\rm q}=0$ mod $2\pi/3$ 
when $\pi/2-\delta(T)<\theta_{\rm I}<3\pi/2+\delta(T)$, 
where $\delta(T)$ is given in \eqref{delta}. 
For the case of $\theta_{\rm I}=0$, 
the RW phase transition is first order at the endpoint in 
the present PNJL calculation. This is consistent with 
the latest LQCD data~\cite{D'Elia-3}. 
In a forthcoming paper, we will analyze the relation 
between imaginary and real $\theta_{\rm I}$. 

\noindent
\begin{acknowledgments}

Authors thank M. Matsuzaki and K. Kashiwa for useful discussions. 
H. K. also thanks M. Imachi, H. Yoneyama and M. Tachibana for useful discussions. 
Y. S. is supported by JSPS Research Fellow.

\end{acknowledgments}


\end{document}